%% file: icml.tex
\documentclass{article}

\usepackage{microtype}
\usepackage{graphicx}
\usepackage{subcaption}
\usepackage{booktabs} 
\usepackage{placeins}

\usepackage{hyperref}

\usepackage[accepted]{icml2025}

\usepackage{amsmath}
\usepackage{amssymb}
\usepackage{mathtools}
\usepackage{amsthm}

\input{math_commands}

\usepackage{url}            
\usepackage{booktabs}       
\usepackage{amsfonts}       
\usepackage{nicefrac}       
\usepackage{microtype}      
\usepackage{xcolor}         
\usepackage{amsmath}
\usepackage{mathtools}
\usepackage{amsthm}
\usepackage{amsmath}
\usepackage{amssymb}
\usepackage{bbm}
\usepackage{tikz} 
\usepackage{wrapfig}
\usepackage{enumitem}

\usepackage[capitalize,noabbrev]{cleveref}

\theoremstyle{plain}
\newtheorem{theorem}{Theorem}[section]

\theoremstyle{definition}

\theoremstyle{remark}

\DeclareMathOperator*{\argmax}{arg\,max}

\usepackage[textsize=tiny]{todonotes}

\icmltitlerunning{When Should We Orchestrate Multiple Agents?}

\begin{document}

\twocolumn[
\icmltitle{When Should We Orchestrate Multiple Agents?}



\icmlsetsymbol{equal}{*}

\begin{icmlauthorlist}
\icmlauthor{Umang Bhatt}{equal,nyu,ati}
\icmlauthor{Sanyam Kapoor}{equal,nyu}
\icmlauthor{Mihir Upadhyay}{nyu}
\icmlauthor{Ilia Sucholutsky}{nyu}
\icmlauthor{Francesco Quinzan}{ox}
\icmlauthor{Katherine M. Collins}{cam}
\icmlauthor{Adrian Weller}{ati,cam}
\icmlauthor{Andrew Gordon Wilson}{nyu}
\icmlauthor{Muhammad Bilal Zafar}{rub}
\end{icmlauthorlist}

\icmlaffiliation{nyu}{New York University, New York, USA}
\icmlaffiliation{ati}{The Alan Turing Institute, London, UK}
\icmlaffiliation{ox}{University of Oxford, Oxford, UK}
\icmlaffiliation{cam}{University of Cambridge, Cambridge, UK}
\icmlaffiliation{rub}{Ruhr University Bochum, Bochum, Germany}

\icmlcorrespondingauthor{Umang Bhatt}{umangbhatt@nyu.edu}

\icmlkeywords{Machine Learning, ICML}

\vskip 0.3in
]



\printAffiliationsAndNotice{\icmlEqualContribution} 

\begin{abstract}
Strategies for orchestrating the interactions
between multiple agents, both human and artificial, can wildly overestimate performance and underestimate the cost of orchestration. 
We design a framework to orchestrate agents under realistic conditions, such as inference costs or availability constraints.
We show theoretically that orchestration is only effective if there are performance or cost differentials between agents. 
We then empirically demonstrate how orchestration between multiple agents can be helpful for selecting agents in a simulated environment,  picking a learning strategy in the infamous Rogers' Paradox from social science, and outsourcing tasks to other agents during a question-answer task in a user study. 
\end{abstract}

\section{Introduction}
The world is full of agents, making decisions and taking actions based on those decisions. Historically, ``agents'' have been restricted to humans: human decision makers, humans assisting other human decision makers, and organizations or institutions taking collective action on behalf of other humans. Increasingly, however, the purview of ``agents'' is expanding to include artificial systems, e.g., large language models (LLMs) or artificial intelligence (AI) agents, including research assistants for structured output generation and computational thought partners for plan execution~\cite{collins2024building}. For a given task, each agent comes with constraints, expertise, and preferences. As AI agents are endowed with exceptional capabilities~\cite{jaech2024openai}, invoking the right agent at the right time is crucial to realizing the potential of AI: the safe and secure use of AI will be under increased scrutiny in the coming years. 
``Orchestration'' has become a common way to route tasks between a set of agents~\cite{fugener2022cognitive,holstein2023human,rasal2024navigating}. In practice, this amounts to selecting an agent with the highest expected utility or choosing the most accurate agent based on offline data.

Existing research around agent orchestration seldom accounts for real-world costs and constraints formally~\cite{keswani2021towards,lai2022human}.
Realistically, some agents may be unsuitable (e.g., lacking expertise,),  unavailable (e.g., no security clearance, misaligned with regulation, preoccupied with another task), or costly (e.g., astronomical inference-time compute costs, high billable rates, large carbon emissions). Failing to consider agent capabilities alongside costs and constraints can overestimate the effectiveness of orchestration. 

Orchestration is not only about the impact on a single agent or a set of agents' ability to achieve a goal. How we orchestrate agents with one another (humans and AI systems) can have ripple effects across networks of interacting agents. For instance, \citet{collins2025revisiting} advocate for AI researchers to consider engaging with the classic social science challenge, Rogers' Paradox~\cite{rogers1988does}, to better understand -- at a coarse, abstract network level -- the impact of human-AI interaction on a population of agents. In Rogers' Paradox \citep{rogers1988does}, agents must adapt either via individual learning or via learning from other agents. Each agent needs to select a learning strategy at every time step: orchestrating when to pick each learning strategy is key.


In this paper, we consider how to orchestrate between agents when the utility, constraints, and costs of each agent are known.
We discuss when non-random orchestration of a given set of agents is worthwhile. 
We take practical steps to understand when orchestrating agents is helpful and to incorporate real world considerations into orchestration mechanisms.
Our contributions are as follows:
\begin{enumerate}
    \item In \Cref{sec:framework}, we lay out a framework that allows for learning agent capabilities online and orchestrates between multiple agents under real-world considerations (e.g. cost, constraints, capability requirements).
    \item  We describe the conditions under which orchestration of agents is sensible by proposing a metric, the \textit{appropriateness} of orchestration (\Cref{sec:simorch}). We show how differences in agent capabilities can affect the effectiveness of orchestration.
    \item  We use our orchestration framework to resolve a variant of Rogers' Paradox in \Cref{sec:RP} by helping agents narrow down the learning strategies worth prioritizing.
    \item We close with a human subject experiment in \Cref{sec:real_orchestration} and find that users are poor decision-makers when allowed to select from a set of assisting agents. We show that orchestration improves user performance.
\end{enumerate}


\section{A Framework for Orchestration}
\label{sec:framework}
We start by describing a framework for orchestrating between a set of agents. While many have attempted to approximate agent capability over the entire data distribution~\cite{gao2021human,charusaie2022sample,hemmer2022forming}, we consider modeling an agent's performance over pre-specified regions that partition the data distribution.
\paragraph{Regions and Agents.}
We denote with $\mathcal{D}$ the input dataset. We posit that the input data distribution can be partitioned into mutually-exclusive regions $\mathcal{R}_1, \dots, \mathcal{R}_M$. For instance, in CIFAR-10 \citep{krizhevsky2009learning} regions correspond to data points with different labels, whereas in MMLU \citep{Hendrycks2020MeasuringMM} they correspond to different subjects, e.g. mathematics, philosophy, astronomy. We use the notation $\mathbb{P}(\mathcal{R}_m)$ to denote the probability of observing a data sample from region $\mathcal{R}_m$. In this framework, the dataset $\mathcal{D}$ consists of triplets $\mathcal{D} = \{ x_t,y_t,r_t \}_{t=1}^N$, where $x_t \in \mathcal{X}$ is the input, $y_t \in \mathcal{Y}$ is the corresponding input label, and $r_t \in \{1,\dots,M\}$ is the region label, indicating if $x_t$ belongs to a region $\mathcal{R}_{r_t}$.

Let $\mathcal{A} = \{A_1, \dots, A_K\}$ be a predefined set of agents, which we desire to orchestrate. 
Every agent is a function ${A_k: \mathcal{X} \to \mathcal{Y}}$, that maps the inputs $x \in \mathcal{X}$ to outputs $y \in \mathcal{Y}$. For example, in the context of Human + Large Language Model (LLM) interaction, a possible set of agents is ${\mathcal{A} = \{ \texttt{HUMAN}, \texttt{AI},  \texttt{HUMAN+AI} \}}$, where $\texttt{HUMAN}$ denotes the decision of an unaided human, $\texttt{AI}$ that of a LLM, and \texttt{HUMAN+AI} denotes that of a human with LLM assistance. For each agent, we denote by $\mathbb{P}(A_k \mid \mathcal{R}_m)$ the probability of correctness of an agent $A_k$ in region $\mathcal{R}_m$, i.e., the fraction of input data such that ${A_k(x_t) = y_t}$.
%

\paragraph{Orchestrating with Agents.}
Given inputs from dataset $\mathcal{D}$ \emph{streamed} one at a time, orchestration consists of selecting the optimal agent $A_k \in \mathcal{A}$ to maximize  \emph{correctness} over the complete dataset. Notably, orchestration is different from the batch setting, where all data points are observed at once.
%
%

We define the \emph{long-running correctness} of an agent $A_k$ as:
\begin{align*}
\begin{split}
\mathsf{C}(A_k) 
&= \sum_{m=1}^M \mathbb{P}(\mathcal{R}_m) \mathbb{P}(A_k \mid \mathcal{R}_m).
\end{split}
\end{align*}
%
Next, we define the \emph{onwards correctness} of an agent as at each step $t \in \{1,\dots,N\}$ of the data stream, which we denote with $\mathsf{C}_{\geq t}(A_k)$. This quantity can be conveniently decomposed into the correctness at time step $t$ $\mathsf{C}_{>t}(A_k)$ and the future long-running correctness $\mathsf{C}_{>t}(A_k)$. Given an observed data point $\{x_t,y_t,r_t\}$ at a fixed time step $t$, the correctness at time step $t$ can be computed as $\mathsf{C}_t(A_k) = \mathbb{P}(A_k \mid \mathcal{R}_{r_t}) $. On the other hand, the future long-running correctness $\mathsf{C}_{>t}(A_k)$ can be computed as ${\mathsf{C}_{>t}(A_k) = \mathsf{C}(A_k)}$, since the future is independent of the current time-step. Then, the onwards correctness of an agent is defined as
\begin{align}
\begin{split}
\mathsf{C}_{\geq t}(A_k) &= \mathsf{C}_t(A_k) \cdot \mathsf{C}_{>t}(A_k), \\
&= \mathbb{P}(A_k \mid \mathcal{R}_{r_t}) \cdot \sum_{m=1}^M \mathbb{P}(\mathcal{R}_m) \mathbb{P}(A_k \mid \mathcal{R}_m).
\end{split}\label{eq:utility}
\end{align}
Intuitively, $\mathsf{C}_{\geq t}(A_k)$ provides us the \emph{utility} of an selecting an agent $A_k$, which acts as a lookahead for the future probability of correctness $\mathsf{C}_{> t}(A_k)$ weighed by the correctness for the given input $x_t$ from region $ \mathcal{R}_{r_t}$. 

The goal of orchestrating with agents consists of selecting an agent $A_{k^\star}$ at each time step that provides the maximum onward utility, i.e.,
$k^\star = \argmax_{k \in \{1,\dots,K\}} ~~ \mathsf{C}_{\geq t}(A_k)$.

\subsection{Estimating Correctness}
\label{sec:orch_estimation}

In practice, both the region probability $\mathbb{P}(\mathcal{R}_m)$ and the agent correctness per region $\mathbb{P}(A_k \mid \mathcal{R}_m)$ are unknown quantities. Therefore, we propose a simple model which allows us to estimate these quantities online as we observe the stream during orchestration using probabilistic inference.

We use the notation $\mathbb{P}(\mathcal{R}_m) \coloneqq w_m$ to denote the probability of observing a region, such that the vector $\boldsymbol{w} = \{w_1,\dots,w_M\}$ represents all such probabilities. 
Denoting the observed stream of data before time step $t$ as $\mathcal{D}_{<t}$, we aim to estimate ${\mathbb{P}(\boldsymbol{w} \mid \mathcal{D}_{<t})}$. 
This quantity is formally referred to as the posterior over $\boldsymbol{w}$ and can be decomposed into a product of a likelihood ${\mathbb{P}(\mathcal{D}_{<t} \mid \boldsymbol{w})}$ and a prior $\mathbb{P}(\boldsymbol{w})$ as prescribed by the Bayes' theorem. 
For categorical variables, like those of region labels, the typical choice is to specify a multinomial likelihood and Dirichlet prior as 
\begin{equation*}
\mathbb{P}(\mathcal{D}_{<t} \mid \boldsymbol{w}) \propto \prod _{m=1}^M w_j^{n_{<t, m}} \ \ \text{and} \ \ \mathbb{P}(\boldsymbol{w}) \propto \prod_{m=1}^M w_m^{\alpha_m - 1},
\end{equation*}
where $n_{<t,m}$ is the number of observed data points from region $\mathcal{R}_m$ until time $t$, and $\alpha_m > 0$ are parameters of the Dirichlet prior. Intuitively, $\alpha_m$ represent a pseudo-count for the corresponding regions $\mathcal{R}_m$, and they allows us to specify our initial belief over the region distribution. Following \citet{murphy2022probabilistic}, this choice conveniently allows for the posterior to also be a Dirichlet distribution as
\begin{equation*}
\mathbb{P}(\boldsymbol{w} \mid \mathcal{D}_{<t}) \propto \prod_{m=1}^M w_m^{(n_{<t,m} + \alpha_m - 1)}.
\label{eq:region_post}
\end{equation*}
Finally, we select a $\boldsymbol{w}_t = \{w_{t,1}, \dots, w_{t,M} \}$ that maximizes the probability under the posterior, which has closed form for a Dirichlet distribution as
\begin{align}
\begin{split}
\boldsymbol{w}_t &= \argmax_{\boldsymbol{w}} ~~ \mathbb{P}(\boldsymbol{w} \mid \mathcal{D}_{<t}), \\
\mathrm{where  }~~ w_{t,m} &= \frac{n_{<t,m} + \alpha_m - 1}{\sum_{j=1}^M (n_{<t,j} + \alpha_j - 1) }.
\end{split}\label{eq:region_map}
\end{align}

In similar vein, we denote the agent correctness per region by $\mathbb{P}(A_k \mid \mathcal{R}_m) \coloneqq c_{km}$ for a fixed agent $A_K$ and region $\mathcal{R}_m$. We aim to estimate the posterior ${\mathbb{P}(\boldsymbol{c}_{km} \mid \mathcal{D}_{<t})}$, where $\boldsymbol{c}_{km} = [1-c_{km}, c_{km}] $ is the vector of probabilities of being correct $c_{km}$ and incorrect $1-c_{km}$. As a special case of the model described above \citep{murphy2022probabilistic}, we arrive at a Beta-Binomial posterior as
\begin{equation*}
\mathbb{P}(\boldsymbol{c}_{km} \mid \mathcal{D}_{<t})
\propto (1 - c_{km})^{(n_{<t,0} + \alpha_{0} - 1)} (c_{km})^{(n_{<t,1} + \alpha_{1} - 1)}
\end{equation*}
where $n_{<t,0}$ and $n_{<t,1}$ represent the number of incorrect and correct responses respectively from agent $A_k$ in region $\mathcal{R}_m$ until time $t$ . Here, $\alpha_0$ and $\alpha_1$ represent the pseudo-counts for the corresponding prior. Selecting $\boldsymbol{c}_{t,km}$ for the highest likelihood, we arrive at the formula
\begin{align}
\begin{split}
\boldsymbol{c}_{t,km} &= \argmax_{\boldsymbol{c}_{km}} ~~ \mathbb{P}(\boldsymbol{c}_{km} \mid \mathcal{D}_{<t}), \\
\mathrm{where  }~~ c_{t,km} &= \frac{n_{<t,1} + \alpha_1 - 1}{(n_{<t,0} + n_{<t,1}) + (\alpha_0 + \alpha_1) - 2 }.
\end{split} \label{eq:actor_region_map}
\end{align}

%

Finally, substituting \cref{eq:region_map,eq:actor_region_map} in our utility from \cref{eq:utility} gives us an \emph{empirical} estimate of the utility of an agent $A_k$ at any given time step $\widehat{\mathsf{C}}_{\geq t}(A_k)$ as,
\begin{equation}
\widehat{\mathsf{C}}_{\geq t}(A_k) = \boldsymbol{c}_{t,k r_t} \cdot \sum_{m=1}^M w_{t,m} \cdot  \boldsymbol{c}_{t,km}. \label{eq:perf_utility}
\end{equation}

\subsection{Practical Considerations}
\paragraph{Cost of Orchestration.}
In the real world, orchestrating with agents incurs costs. For instance, API providers charge by the number of tokens in both the input and output of an API request to an LLM, such that the cost per input may be variable \citep{jaech2024openai,claude35}.
We estimate the cost of an agent offline before orchestration using held-out data from all regions, and denote $\gamma_{km}$ as the cost of agent $A_k$ in region $\mathcal{R}_m$.

Unlike prior work \cite{gao2021human}, we account for this cost, by incorporating it directly with the empirical utility derived in \cref{eq:perf_utility}. The \emph{total empirical utility} is defined as a ratio for a given input from region $\mathcal{R}_{r_t}$, as
\begin{equation*}
\widehat{\mathsf{U}}_{\geq t}(A_k) = \frac{\widehat{\mathsf{C}}_{\geq t}(A_k)}{\gamma_{kr_t}}. 
\end{equation*}
We then proceed with orchestration by selecting the agent $A_{k^{\star}}$ with the highest total empirical utility,
\begin{align*}
k^\star \in \argmax_{k \in \{1,\dots,K\}} ~~ \widehat{\mathsf{U}}_{\geq t}(A_k).
\end{align*}
\paragraph{Additional Constraints on Orchestration.} 
Other than the costs described in the previous section, there may be several additional constraints.
For instance, some agents may be unable to operate on data from some regions, while others may be frequently unavailable due to their high demand, and others may perform inference too slowly. 
Such constraints can be framed as a Boolean mask atop the $K$ agents at each time step $t$. Most work on orchestrating of agents, especially those routing between a human and an AI system, assume that all agents $\gA$ are available at every time step~\cite{gao2021human,bhatt2023learning}. In reality, there may be fewer feasible agents.

We represent each constraint as $g_t: \gX \times \gA \to \{0,1\}^{K}$, where $t$ is the current timestamp. The output $g_t(x_t; A_k)$ tells us if agent $A_k$ is feasible for a task $x_t$. To justify and illustrate the practical use of these constraints, 
consider the EU AI Act \cite{EUAIAct}. The EU AI Act stipulates that humans must be in the loop for any high-risk algorithmic decision-making system.
A simplistic constraint for EU AI Act compliance, i.e., the feasibility of an agent, may be written as
\begin{align*}
g_t(x_t; A_k) \coloneqq & 
\left \{
\begin{array}{ll}
1 & \text{if $x_t$ is not high risk or $A_k$ is human};\\
0 & \text{otherwise}
\end{array}
\right .
\end{align*}

When we proceed with orchestration under constraints, we can simply select the feasible agent with the highest total empirical utility, as
\begin{align*}
k^\star \in \argmax_{k \in \{1,\dots,K\}\colon g_t(x_t; A_k) = 1} ~~ \widehat{\mathsf{U}}_{\geq t}(A_k).
\end{align*}
It is worth noting, however, that this simplified utility does not account for future constraints which we leave for interesting followup work. 

\section{Simulating Orchestration}
\label{sec:simorch}
Having setup the basic framework for orchestration in \cref{sec:framework}, we now move to understanding various scenarios under which orchestration is worthwhile. The goal of agent orchestration is to know which agents are suitable (i.e., capable to perform the assigned task well), feasible (e.g., available), and economical (i.e., cost-effective).

\subsection{Appropriateness of Orchestration}
\label{sec:approp}
Mere availability of multiple agents does not imply that we will benefit from orchestration. 
For instance, consider the trivial case where all agents perform equally well on all regions, i.e. the empirical utility in \cref{eq:perf_utility} is approximately equal for all agents.
In this scenario, selecting any agent to serve all inputs in the data stream, or even selecting an agent uniformly at random for every input will not lead to an improved correctness on the dataset with the orchestrator.

Therefore, to assess whether orchestration is worthwhile, we propose to estimate the \emph{appropriateness} of orchestration as the ratio of the theoretical maximum achievable correctness $\mathsf{C}_{\max}$ to the expected correctness achievable under a random orchestrator $\mathsf{C}_{\mathrm{rand}}$ (i.e. an orchestrator that selects agents at random at each time step) as,
\begin{align}
\mathsf{App} =\frac{\mathsf{C}_{\max}}{\mathsf{C}_{\mathrm{rand}}}. \label{eq:approp}
\end{align}
We assume that all agents have a non-zero correctness on all regions, which is largely the case in practice and allows \cref{eq:approp} to be a well-defined quantity.
In this definition, the theoretical maximum $\mathsf{C}_{\max}$ is achieved by an orchestrator which selects the agent corresponding to maximum correctness in input region $r_t$, i.e. ${\argmax_k ~~ \mathbb{E}[\mathbb{P}(A_k \mid \mathcal{R}_{r_t})]}$.

Next, we define the \emph{dissimilarity} between two agents as,
\begin{equation*}
d(A_k, A_h) = \max_m ~~ \exp \left|\log \frac{\mathbb{P}(A_k \mid \mathcal{R}_m)}{\mathbb{P}(A_h \mid \mathcal{R}_m)} \right|.
\end{equation*}
This metric essentially tells us if two agents are ``similar" (when it's close to $1$) or ``dissimilar" (when it's greater than one). We can say that an agent $A_k$ is redundant if there exists another agent $A_h $ such that $d(A_k, A_h) =  1$. Next, we provide a lower bound for the appropriateness of the orchestration given a set of agents.
\begin{theorem}
\label{lemma:approp}
Let $A_{\mathrm{rand}}$ denote an agent chosen uniformly at random. For every $\varepsilon, \delta \in (0,1)$ there exists an orchestration problem such that an agent $A_i $ chosen uniformly at random yields 
\begin{equation*}
    \frac{\mathsf{C}_{\max}}{\mathsf{C}(A_{\mathrm{rand}})} \geq \min_{k,h \colon \mathsf{C}(A_k) \neq \mathsf{C}(A_h)}d(A_k, A_h) \geq \frac{1}{1-\varepsilon},
\end{equation*}
 with probability at least $1-\delta$. In particular, it holds
\begin{equation*}
    \frac{\mathsf{C}_{\max}}{\mathsf{C}_\mathrm{rand}} \gtrsim \frac{1}{1-\varepsilon},
\end{equation*}
 for $\delta \to 0$.\footnote{This statement formally means that
    $\lim_{\delta \to 0}\frac{\mathsf{C}_{\mathrm{max}}}{\mathsf{C}_{\mathrm{rand}}} \geq \frac{1}{1-\varepsilon}$.}
\end{theorem}
The proof of Theorem \ref{lemma:approp} is deferred to Appendix \ref{appendix:proof}. Intuitively, the proof of Theorem \ref{lemma:approp} consists of constructing an example of orchestration, for which the claim holds.
%

\cref{lemma:approp} helps us reason about the appropriateness of orchestration in a principled manner. We immediately see how the appropriateness of orchestration is closely connected to the similarity of two agents via a lower bound. A higher similarity qualitatively implies that orchestration is not worthwhile, such that the appropriateness of orchestration is trivially equal to the lower bound with value $1$. To the contrary, as the agents start varying in correctness, we see greater value for orchestration as one would expect.

\subsection{Understanding Orchestration Scenarios}
\label{sec:understanding_orch}

It is instructive to understand
how orchestration behaves in various scenarios.
In this section, we construct a synthetic dataset by hand-coding various instances of the correctness probability of an agents in a region $\mathbb{P}(A_k \mid \mathcal{R}_m)$ for $K= 4$ agents and $M=3$ regions, while using a uniform probability over regions.  One can easily extend the setup to account for constraints by limiting to only the feasible subset of agents.
But to simplify exposition, we do not consider constraints in what follows.

All agents can have different ``expertise'' across multiple regions. For instance, one agent may be achieve higher correctness in the topics related to mathematics, whereas another in the topics related to philosophy. Based on a user study, \citet{bhatt2023learning} identify three expertise profiles (illustrated in \cref{fig:expertise_profiles}) which we analyze below.

\begin{enumerate}[label=(\roman*)]

\item \textbf{Approximately Invariant.} Under this scenario, all agents are approximately equal in correctness over all regions. More precisely, ${\mathbb{P}(A_k \mid \mathcal{R}_m) \approx \mathbb{P}(A_h \mid \mathcal{R}_m)}$ for all $k,h \in \{1,\dots,K\}$ and for all $m \in \{1,\dots,M\}$.

In this case, we immediately see that the appropriateness of orchestration approximately equals one, since ${\mathsf{C}_{\mathrm{max}} \approx \mathsf{C}_{\mathrm{rand}}}$ with probability (w.p.) one. This expertise profile achieves the lower bound shown by \cref{lemma:approp}, and therefore is the least appropriate for orchestration as one would intuitively identity.
Notably, even when we account for the cost associated with employing any available agent, the overall correctness after orchestration still remains similar.
As a consequence, such a scenario remains the least amenable to orchestration as noted by appropriateness close to one in \cref{fig:profiles_approp}.

\item \textbf{Dominant.} Here, we have one agent $A_k$ which is strictly better than any other agent $A_h$ over all regions, i.e. ${\mathbb{P}(A_k \mid \mathcal{R}_m) > \mathbb{P}(A_h \mid \mathcal{R}_m)}$ for all $m \in \{1,\dots,M\}$ and all agents $A_h$ such that $k\neq h$.

The appropriateness of orchestration in this case is apparent. 
The theoretical maximum correctness $\mathsf{C}_{\mathrm{max}}$ can be achieved by always selecting the dominating agent, e.g. $A_1$ in \cref{fig:expertise_profiles}(ii), and corresponds to its expected correctness. 
However, the expected correctness under a random orchestrator will be weighted down by the sub-optimal correctness of the remainder of the agents. Therefore, we will certainly achieve appropriateness greater than the lower bound of $1$.

When accounting for the cost, two cases are noteworthy --- (a) the cost for the dominant agent is strictly lower than other agents which only exaggerates the utility $\mathsf{U}_{\geq t}$ of the dominant agent while preserving the appropriateness of orchestration, and (b) the cost is misaligned against the dominant agent such that the utility of the agents can change in unexpected ways, often in the worst case completely destroying the appropriateness of orchestration as we see in \cref{fig:profiles_approp}. Case (b) can arise in scenarios when a sub-optimal agent provides overly favorable cost.

\begin{figure}[!t]
    \centering
     \includegraphics[width=.9\linewidth]{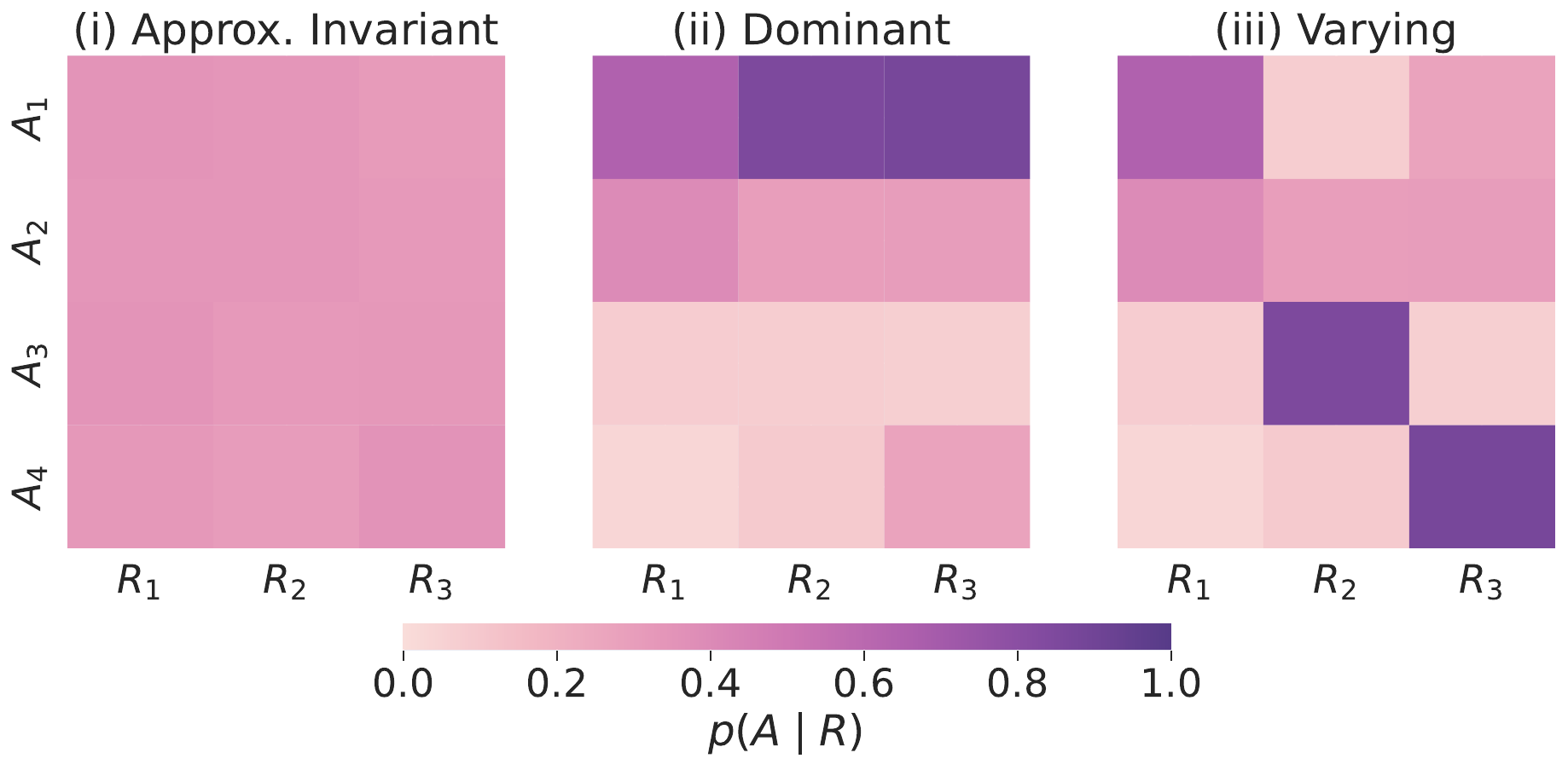}
     \vspace{-0.15in}
    \caption{Using an example with $K=4$ agents and $M=3$ regions, we visualize how orchestration can be divided into three qualitative scenarios in terms of correctness ${\mathbb{P}(A_k \mid \mathcal{R}_m)}$ --- (i) \textbf{Approximately Invariant} where all agents are approximately equal in correctness over all regions, (ii) \textbf{Dominant} where a single agent is strictly better on all regions than any other agent, and (iii) \textbf{Varying} where agents may occupy different regions of high correctness. See \cref{sec:understanding_orch} for discussion.}
    \label{fig:expertise_profiles}
\end{figure}

\item \textbf{Varying.} In this scenario, the expertise of agents may vary across different regions. i.e. there exist agents $A_k$ and $A_h$, and regions $\mathcal{R}_m$ and $\mathcal{R}_\ell$, such that ${\mathbb{P}(A_k \mid \mathcal{R}_m) > \mathbb{P}(A_h \mid \mathcal{R}_m)}$ and ${\mathbb{P}(A_h \mid \mathcal{R}_\ell ) > \mathbb{P}(A_k \mid \mathcal{R}_\ell)}$.

This scenario involves complex interactions between the performance of different agents across regions.
The theoretical maximum can be achieved by selecting the most correct agent for every corresponding input region.
However, the performance may drop as a consequence of selecting agents randomly to compute $\mathsf{C}_{\mathrm{rand}}$ will depend upon the maximum performance difference between two agents $d(A_k,A_h)$ as formally noted in \cref{lemma:approp}, thus varying the appropriateness of orchestration.
Further, as we increase the number of redundant agents with similar capabilities across regions, the appropriateness decreases.

When accounting for the cost of orchestration, a few cases are of interest --- (a) when the most correct agents in a region tend to be disproportionately higher cost, the orchestration can fail as we see in \cref{fig:profiles_approp}, and (b) even when one agent has a strictly lower cost than all other agents, orchestration may be rendered less effective than desired.

\end{enumerate}

\begin{figure}[!t]
    \centering
    \includegraphics[width=.85\linewidth]{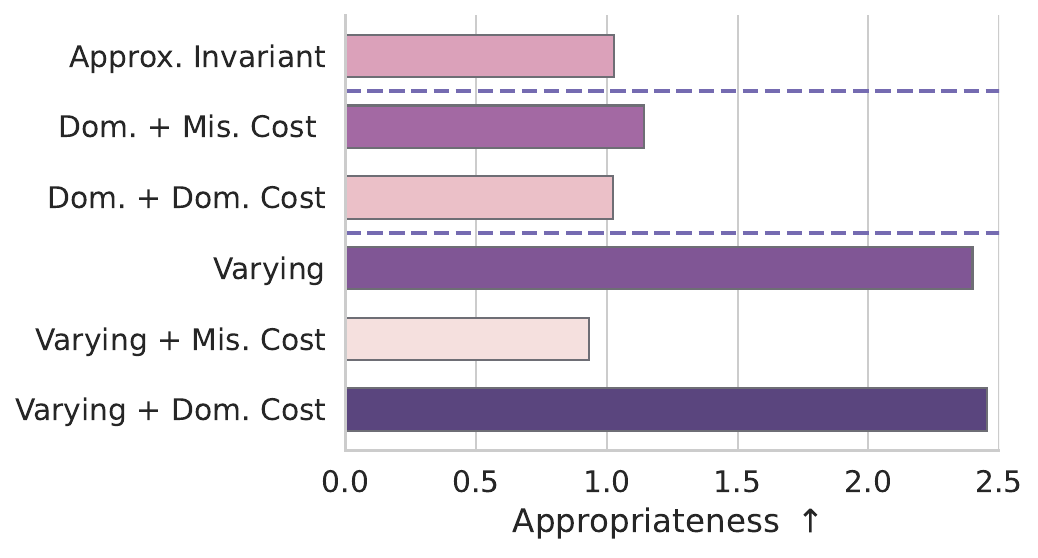}
    \vspace{-0.15in}
    \caption{We compute the appropriateness (\cref{eq:approp}, higher is better) for select cases of expertise scenarios in \cref{fig:expertise_profiles}, accounting for cost in addition. 
    An approximately invariant profile stays least appropriate for orchestration. A dominant (\texttt{Dom.}) profile with dominant cost can be less appropriate for orchestration than a profile with misaligned (\texttt{Mis.}) costs where careful agent selection becomes more important.
    The appropriateness of a varying expertise profile can be hampered by misalignment of costs. 
    Evidently, our measure of appropriateness, while being simple, can capture such nuances for the effectiveness of orchestration. See \cref{sec:understanding_orch} for discussion.}
    \label{fig:profiles_approp}
\end{figure}

We highlight two additional challenges in constrained orchestration.
Firstly, the presence of constraints may prevent the orchestrator from choosing the optimal policy at each time step. For instance, a dominant agent may be unavailable due to usage quota restrictions forcing the usage of a sub-optimal agent. 
Therefore, constrained orchestration leads to \textit{necessarily} worse correctness in such scenarios.
Nevertheless, orchestration can still provide value such that the remainder of the agents behave similar to the varying expertise case.
Secondly, unlike the batch setting where even in the presence of constraints we can devise an optimal plan ahead-of-time, the streaming setting limits our ability to plan ahead for future tasks.

Finally, we note that the choice of priors, i.e. $\alpha$'s for the estimation of the correctness probabilities and region probabilities as described in \cref{sec:orch_estimation} can have significant impact on the effectiness of orchestration. When available, it is always advisable to start from non-uniform prior to accelerate estimation.

\section{Resolving Rogers' Paradox}
\label{sec:RP}

Rogers' Paradox involves a simulation of agents who are trying to survive in a changing world. Agents can learn about the world through one of two means: through costly ``individual learning'' or cheap (but potentially time-lagged) ``social learning'' from others around them. Alan Rogers discovered that simulated populations in this environment with abundantly available cheap social learning are no better off in terms of their collective understanding at equilibrium than when social learning is unavailable and everyone is an individual learner~\citep{rogers1988does}. However, such a finding runs in stark contrast with what we may expect from our understanding of cultural evolution: that part of human thriving arises from ratcheting up on what others have done. 

\begin{figure}[!t]
    \centering
    \includegraphics[width=0.9\linewidth]{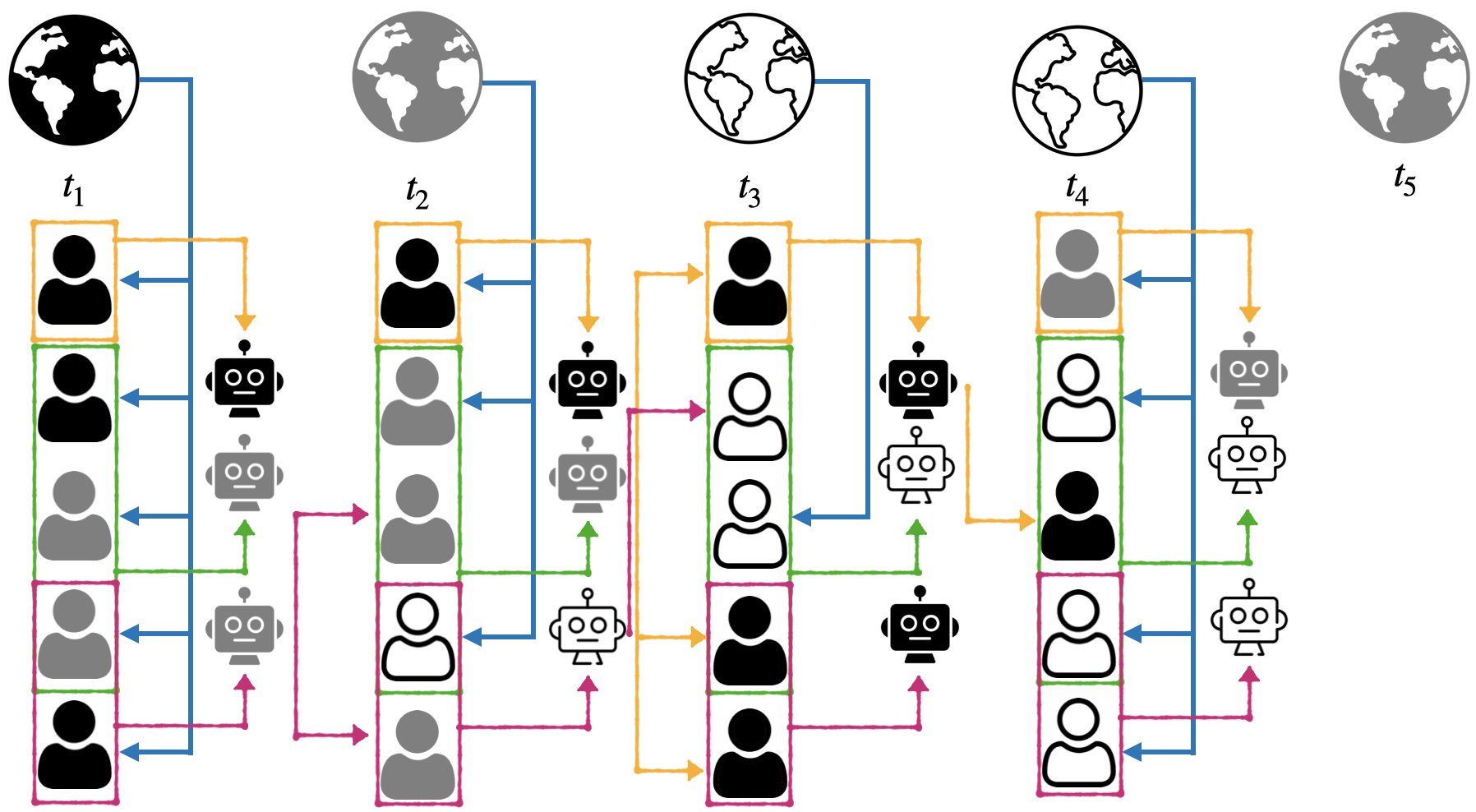}
    \caption{Example abstract setting of Rogers' Paradox whereby humans may choose to learn about the world individually, from another human, or from one of multiple different possible AI systems (which in turn learn from people). People and AI systems' understanding of the world evolves overtime, while the world itself may change. Orchestration provides one way to help humans navigate which agent to learn about the world from at each step.}
    \label{fig:multi_RP}
\end{figure}

Decades of work following this finding have sought to develop strategies that recover the benefits of social learning; however, as \citet{collins2025revisiting} points out -- these simulations are ripe for revisiting when we consider the impact of humans socially learning from AI systems which too may socially learn from us. No work - to our knowledge - has yet looked at \textit{orchestration} of humans with many \textit{different} AI systems in such a network model, which we depict in Figure~\ref{fig:multi_RP}. Can orchestration boost collective population understanding, without falling prey to Rogers' Paradox, when there is not just one but many ``AI'' nodes in such a network? 

In practice, there may be many AI systems from which members of a population can learn.
By introducing multiple AI systems, we represent learning strategies with AI assistance that demonstrate in the real world more faithfully.
However, this raises a computational challenge: we have now introduced multiple learning strategies for a person to pick from at every timestep. A person can now individually learn from the world at a high cost, socially learn from another person, or socially learn from one of a few specialized AI systems that each learn from a different subset of a population. This is where orchestration can help a person to decide how to learn.
We use our orchestration framework to select a teaching style for every learner at every timestep. In our simulations, we show that introducing orchestration greatly increases the population's equilibrium performance and resolves Rogers' Paradox.

Concretely, we replicate the simulations of~\citet{collins2025revisiting} with four extensions: individual differences, availability of multiple AI systems, two types of constraints, and orchestration.\footnote{We caveat that our simulations are coarse and meant to offer general insights into the \textit{collective} societal ripple effects of orchestration; much work is needed to enrich such simulations.} We consider a population of $1000$ simulated human agents who have to adapt to a (stochastically) changing world that has a small probability of changing its state at the end of each timestep. During every timestep, each human agent can individually learn from the world (with normally distributed individual differences in success probability), socially learn from one of the other 999 agents, or socially learn from one of three AI systems (I, II, and III). AI system I is ``trained on'' the outputs of the agents who socially learned on the previous timestep, meaning we set the AI's adaptation level to the mean of these agents. Similarly, AI system II reflects the population mean of the agents who individually learned on the previous timestep, and AI system III is the population mean of all human agents irrespective of their learning strategy on the previous timestep. Each human agent, in principle, has to choose between approximately $1000$ different learning strategies at every time steps in hopes of adapting to an ever-changing world.
As humans may struggle to track the expected value of thousands of different agents, orchestration can lend a helping hand.

In our \textit{baseline} setting, we allow simulated human agents to either learn individually, learn socially from one of the three AI systems chosen at random, or learn socially from another simulated human agent chosen at random. This is effectively learning without orchestration, as agents are not directed toward the type of learning they ought to use. We find that the population reaches an equilibrium adaptation level of $0.578$ which is almost the same as the equilibrium reached by a population consisting exclusively of individual learners  as predicted by Rogers' Paradox.
\begin{figure}[!t]
    \centering
    \includegraphics[width=0.8\linewidth]{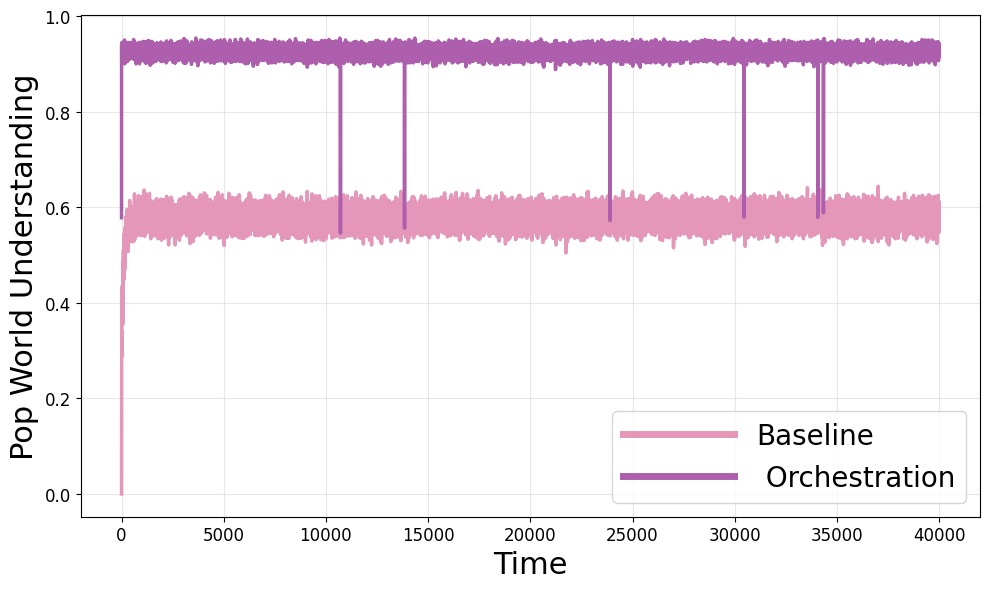}
    \vspace{-0.15in}
    \caption{Comparing the collective world understanding over time in a network of agents that can learn socially from each other or can learn socially from one of three AI systems. The baseline recovers classic Rogers’ Paradox findings; however, orchestration when each learner should adapt to a specific agent within the network resolves paradox to obtain collective world understanding.}
    \label{fig:rogers_solution}
\end{figure}
In our \textit{orchestration} variant, the orchestrator tracks the expected performance of every agent and learning strategy. The world state can be thought of as the different regions that need to be orchestrated over. 
We also introduce two real-world constraints on learning. Our first constraint mandates that each agent can only learn from other nearby agents: that is, agent $i$ can only learn from agents $[i-5,i+5]$.
Our second constraint makes one AI unavailable for learning from in each region (i.e. which AI is unavailable changes every time the world state changes). For a given timestep, the orchestrator needs to solve 1000 constrained orchestration problems by deciding the optimal learning strategy (i.e., how to learn, and who from) for each human agent, subject to the constraints. 

In Figure~\ref{fig:rogers_solution}, we show that orchestration leads to a much higher equilibrium adaptation level for the population of $0.926$, i.e.  
we find that orchestration resolves Rogers' Paradox and ensures that a population actually benefits from the availability of AI systems that enable cheap social learning.


\section{Orchestration with Real Users}
\label{sec:real_orchestration}


Understanding the impacts of outsourcing to agents in the context of orchestrating human decision-making is urgently needed. To that end, we explore such impacts by experimenting with orchestrating when users can \textbf{elect} to outsource decision-making, instead of users seeing AI assistance before making their own decision. More precisely,  in the context of AI assistance, AI agent responses are automatically shown to users and users are free to disagree with the agent's recommendation~\cite{swaroop2024accuracy}; however, here we let users decide whether they want to outsource (to an AI or human agent) and, if the user decides to outsource, they are \textit{locked in} to the agent's decision with no ability to override. We show how orchestration excels in this setup. 

\paragraph{Experimental Setup.}
To explore the effects of orchestration on user decision-making, we run a user study with $80$ participants recruited from the crowdsourcing platform, Prolific. Participants partake in a series of mathematics problems drawn from MMLU~\cite{Hendrycks2020MeasuringMM}. 
We consider multiple choice questions from three categories: Elementary Mathematics, High School Mathematics, College Mathematics. We can view each as its own region per Section~\ref{sec:framework}. 
Before considering orchestration, we run a pilot study on $20$ participants to learn user performance with no assistance: see Appendix~\ref{userstudy_app} for details. In this pilot, we ask participants to answer $60$ questions, $20$ per region, without any decision aid. This gives a performance estimate across the population. We find that, on average, participants are correct around $80\%$ of the time on Elementary Mathematics, $55\%$ High School Mathematics, and $44\%$ on College Mathematics. Random guessing would yield $25\%$. 

\begin{figure}[!t]
    \centering
    \includegraphics[width=.9\linewidth]{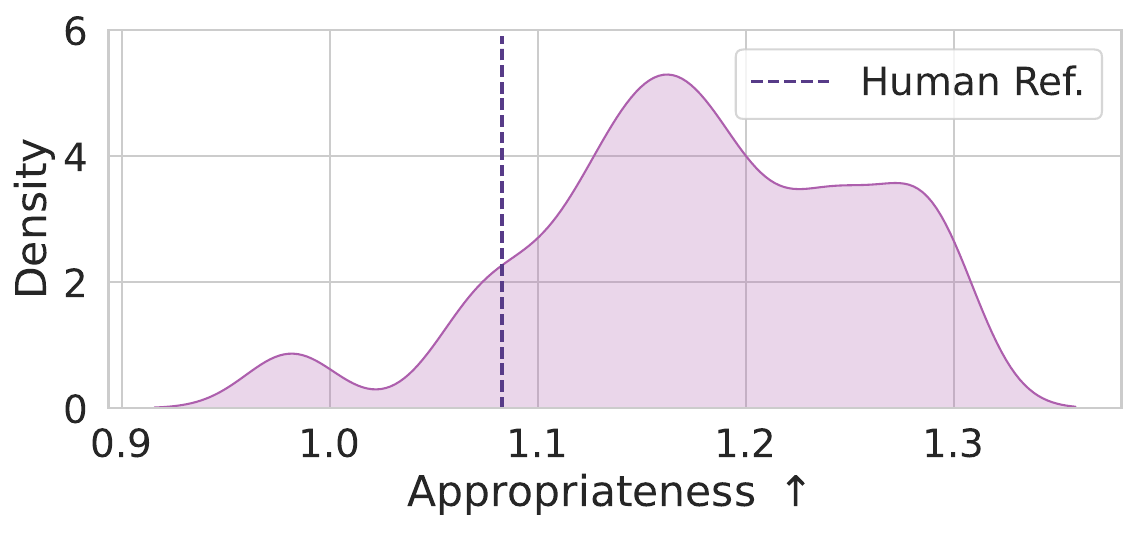}
    \vspace{-0.15in}
    \caption{For the participants across our user study (see \cref{sec:real_orchestration}), we plot the density of appropriateness over the three selected regions of MMLU \citep{Hendrycks2020MeasuringMM} and two agents - one that performs equal to the human population average per region, and another the LLM. The shaded area to the left of human reference corresponds to scenarios where orchestration is less likely to provide benefit, for instance due to strong dominant or approximately invariant profiles for some participants.}
    \label{fig:pilot_approp}
\end{figure}

In our main study, users are permitted to answer questions on their own, or choose to outsource their decision, either to an AI agent or another human.  The AI agent is an LLM, Llama-3-70B~\cite{touvron2023llama}, with $62\%$ accuracy on Elementary Mathematics, $44\%$ High School Mathematics, and $51\%$ on College Mathematics. 
The human agent's decision is assumed to have the accuracy of the average in the population from the pilot above.
During the task, users are awarded $10$ points for each correct answer they provide. 
However, outsourcing incurs a cost: outsourcing to a Human Agent yields costs $7$ points and outsourcing to a AI Agent yields costs $3$ points. This cost structure reflects real-world outsourcing, where the cost of querying an expert is higher than that of an AI agent. future work can explore alternative cost schemes.
We plot the distribution of appropriateness of orchestration for each user in our pilot study in \Cref{fig:pilot_approp}. We find that orchestration may not be advantageous for the fraction of users who have dominant or invariant profiles. 

To assess the impact of orchestration on users (i.e., participants), we consider three experimental conditions:
\begin{enumerate}
    \item \textbf{Baseline}: We let users elect when to outsource to either agent. There is no orchestration. 
    \item \textbf{Orchestration}: We use our orchestration framework to suggest to users which agent to use, if at all. Users are free to ignore our suggestion.
    \item \textbf{Constrained Orchestration}: We again orchestrate suggestions per our framework; however, if the user is wrong on a specific question with no outsourcing, we constrain users and force them to outsource the next question from that region.
\end{enumerate}

\begin{figure}[!t]
    \centering
    \includegraphics[width=0.9\linewidth]{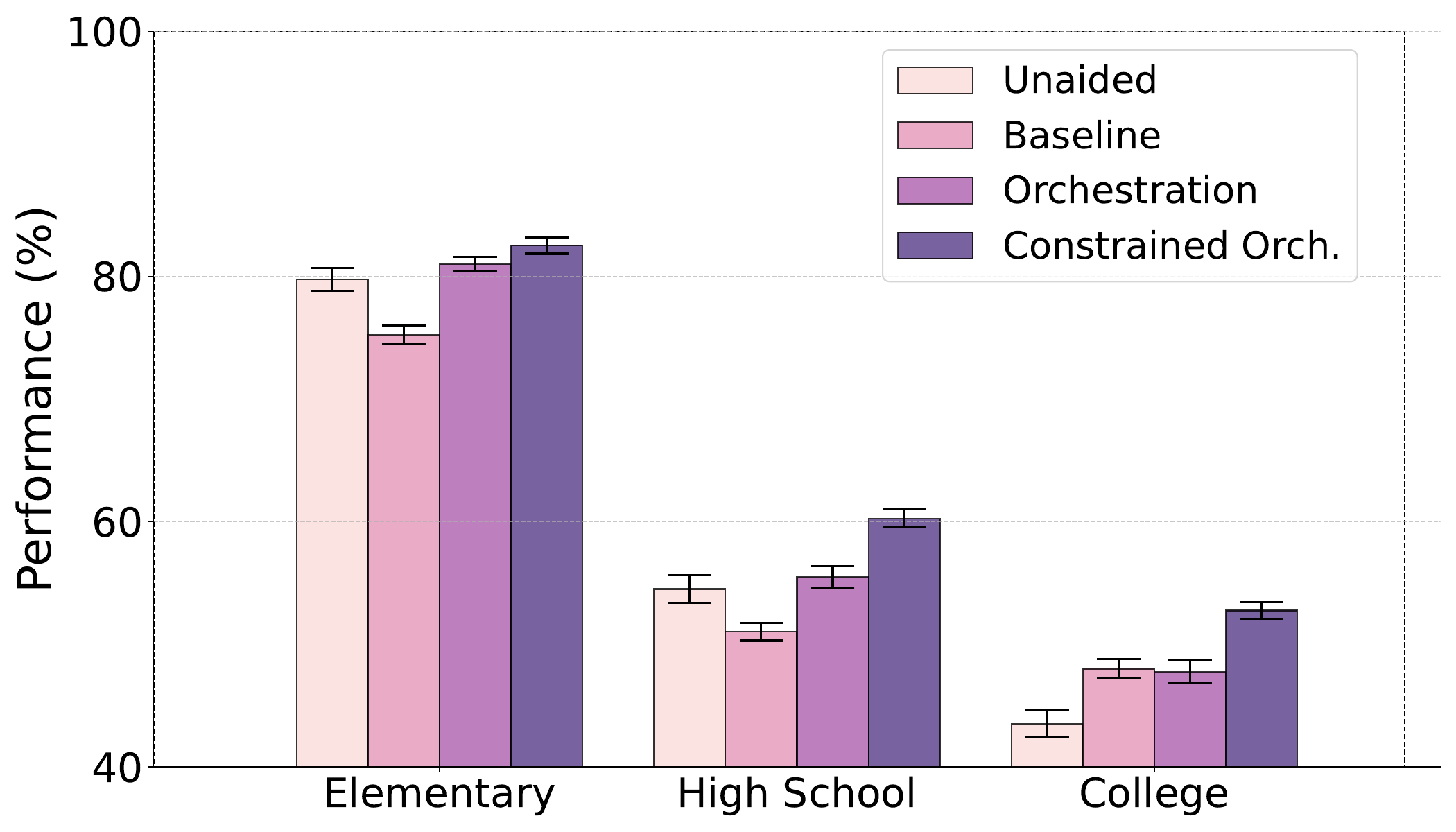}
    \caption{Adding orchestration into the users' workflow improves their decision-making, compared to alternatives. The unaided variant comes from our pilot study where users have no agents to support them. The baseline lets users use the human agent or AI agent without orchestration. When we constrain users after they err on their own, users excel in all regions likely due to increased alterness on the task. These results are expanded in Table~\ref{tab:lockin-performance}.}
    \label{fig:hse}
\end{figure}

\paragraph{Results.} Recall from our pilot that users are on average $60\%$ accurate on the entire task, and struggle with the College mathematics questions getting $44\%$ correct. In the baseline variant, users are surprisingly only $58\%$ accurate: the introduction of decision aids (AI agent and human agent) with no orchestration harms users' decision-making abilities, especially on Elementary and High School Mathematics questions.
When we orchestrate and suggest to users which agent to use, we regain that performance drop, as users are around $61\%$ accurate. Even more promisingly, we find that users achieve $65.2\%$ accuracy when we introduce a constraint, where users must outsource to an agent if they previously made an unaided mistake. In Figure~\ref{fig:hse}, we depict how well users perform by region.

These results are promising. We find that users who have access to agents without orchestration struggle to know when outsourcing to an agent is beneficial. In some sense, users are poor orchestrators when given increased optionality of decision aids but no guidance on when to use which aid. Upon orchestrating agents \textit{on behalf of users} via gentle suggestion, we notice that performance improves. We imagine that altering the affordances we leverage to encourage agent use is a ripe direction for future work: this includes adding nudges or frictions to the interface used to increase agent reliance~\cite{sunstein2014choosing,guo2024decision,collins2024modulating}, altering the underlying orchestration costs based on a user's willingness to pay~\cite{von2022physicians}, or choosing to orchestrate agents without knowing what the suggested agent would be.

In Appendix~\ref{userstudy_app}, we include compare these results to a study with no lock-in: users are permitted to override the agent's response. We find that performance without lock-in improves in the baseline (e.g., access to agents but no suggested orchestration); this suggests that users may use agents as a crutch to refine their own decision-making. Constrained orchestration with no lock-in achieves the best performance on our task. In practice, simulating no lock-in is nontrivial, as it requires high-fidelity human simulators~\cite{tjuatja2024llms}. Future work can extend our framework to more complex rewards signals, where feedback is not immediate, and consider what to do when new agents appear.

\section{Related Work}
Work on orchestrating agents has been studied in various contexts, including delegating between human agents~\cite{hamman2023managing}, choosing between machine learning models~\cite{linhart1986model,riquelme2021scaling}, and routing between combinations of models and humans~\cite{milewski1997delegating,lubars2019ask,lai2022human}.
AI-assisted decision-making has studied this under the guise of deferral~\cite{madras2018predict,mozannar2020consistent,mozannar2022teaching}, delegation~\cite{lai2023towards}, reject-option classifiers~\cite{cortes2018online}, or the imposition of selective frictions~\citep{collins2024modulating}. 

Most common is the dyad setting (One Human and One AI), where tasks are routed between a decision maker and an AI system. These works attempt to orchestrate combinations of humans with AI are or are not useful~\citep{swaroop2024accuracy,vaccaro2024combinations}, or identify settings for complementarity~\citep{bansal2021does,steyvers2022bayesian,gao2023learning}. 
Others have used constraints in routing and planning problems~\cite{nunes2017taxonomy,fuchs2023optimizing}, in resource allocation~\cite{azizi2018designing}, and in matching problems like those assigning teachers to students based on representation alignment~\citep{sucholutsky2024representational}.
There is little work, to the best of our knowledge, that permits multiple support agents (both AI and non-AI) available to decision-makers in the same sequential decision-making task under real world constraints.

\section{Conclusion}
We set out to question when orchestrating a set of agents is worthwhile. We consider a sets of many agents that may include various AI systems, multiple humans, or other hybrid. Our framework permits learning agent capabilities online while accounting for constraints on the task itself. The metric we propose proves useful in knowing if a set of agents can be orchestrated better than random sampling. We show that orchestration can resolve Rogers' Paradox by helping agents select between thousands of learning strategies to maximize collective world understanding. Our user study shows that for a difficult task, providing a set of agents without orchestration can be harmful to user performance: we introduce constrained orchestration to eek out gains.

As AI agents become pervasive, understanding the nuances of orchestrating their use in the real world will be important. 
Formally studying whether orchestration improves upon removing an agent could be a worthwhile pursuit. This may require quantifying the value of each agent to orchestration under non-deterministic constraints. 
If specific agents lack the requisite skills in a certain region (or if no agent is capable enough), upskilling agents is a natural fix. In the end, it is not clear if orchestration is helpful outright. It is key to take a more measured and nuanced approach to deploying a set of agents into the wild.

\section*{Impact Statement}

This paper presents work whose goal is to advance the field of 
human-AI interaction and machine learning. There are many potential societal consequences 
of our work. We are motivated by empowering users to be appropriate and effective in their use of AI systems in practice. We hope that orchestration of AI systems becomes commonplace yet nuanced, as blanket use of AI systems could cause harm.

\section*{Acknowledgements}
UB and FQ acknowledge support by ELSA: European Lighthouse on Secure and Safe AI project
(grant agreement No. 101070617 under UK guarantee). 
Views and opinions expressed are, however, those of the author(s) only and do not necessarily reflect those of any of these funding agencies, European Union or European Commission. AW acknowledges support from a Turing AI Fellowship under grant EP/V025279/1, EPSRC grants EP/V056522/1 and EP/V056883/1, and the Leverhulme Trust via CFI. KMC acknowledges support from a Marshall Scholarship, King's College Cambridge, and the Cambridge Trust. 
MBZ was supported by UA Ruhr Researh Center for Trustworthy Data Science and Security.
We thank Kendall Brogle, Valerie Chen, Guy Davidson, and Brenden Lake for their thoughts and comments on this line of work.

\bibliography{main}
\bibliographystyle{icml2025}

\newpage
\appendix
\onecolumn

\section{Proof of Theorem \ref{lemma:approp}}
\label{appendix:proof}
\begin{proof}
We construct a counterexample for this problem. Let $K = \left \lceil 1/\delta\right \rceil$ be the number of agents we orchestrate. We assume a single region $\mathcal{R}_1$ such that $\mathbb{P}(\mathcal{R}_1) = 1$. Suppose that the agents are such that $\mathbb{P}(A_1 \mid \mathcal{R}_1) = 1$ and $\mathbb{P}(A_k \mid \mathcal{R}_1) = 1 -  \varepsilon $ for all $k \geq 2$. For this example, we show the following claims
\begin{equation}
\label{eq:stronger_claim}
\frac{\mathsf{C}_{\mathrm{max}}}{\mathsf{C}(A_{\mathrm{rand}})} = \min_{k,h \colon \mathsf{C}(A_k) \neq \mathsf{C}(A_h)}d(A_k, A_h) = \frac{1}{1 - \varepsilon} \text{     w.p. at least $1 - \delta$} \text{     and     } \lim_{\delta \to 0}\frac{\mathsf{C}_{\mathrm{max}}}{\mathsf{C}_{\mathrm{rand}}} = \frac{1}{1 - \varepsilon}.
\end{equation}
Note that from \eqref{eq:stronger_claim} the claim follows.

We prove the claim by first computing $\mathsf{C}_{\max}$ and $\mathsf{C}(A_{\mathrm{rand}})$ directly. To compute $\mathsf{C}(A_{\mathrm{rand}})$, note that the optimal solution is to always choose agent $A_1$. Hence, it holds $\mathsf{C}_{\max} = \mathsf{C}(A_1) = 1$. In order to compute $\mathsf{C}(A_{\mathrm{rand}})$, note that selecting an agent $A_k$ with $k \geq 2$ yields $\mathsf{C}(A_k) = 1 - \varepsilon$. Since there are $K = \left \lceil 1/\delta\right \rceil $, then w.p. at least $(1 - \delta) $ it holds $\mathsf{C}(A_{\mathrm{rand}}) = 1 - \varepsilon$. Putting things together, it holds
\begin{equation}
\label{eq:123}
\frac{\mathsf{C}_{\mathrm{max}}}{\mathsf{C}(A_{\mathrm{rand}})} = \frac{1}{1 - \varepsilon},
\end{equation}
with probability at least $1 - \delta$. To conclude the proof, since $\mathsf{C}(A_2) = \dots = \mathsf{C}(A_K)$, we have that
\begin{equation}
\label{eq:456}
\min_{k,h \colon \mathsf{C}(A_k) \neq \mathsf{C}(A_h)}d(A_k, A_h) = d(A_1, A_2) = \exp \left|\log \frac{\mathbb{P}(A_1 \mid \mathcal{R}_1)}{\mathbb{P}(A_2 \mid \mathcal{R}_1)} \right| = \frac{1}{1 - \varepsilon}.
\end{equation}
By combining \Eqref{eq:123} with \Eqref{eq:456} it holds
\begin{equation*}
\frac{\mathsf{C}_{\mathrm{max}}}{\mathsf{C}(A_{\mathrm{rand}})} = \min_{k,h \colon \mathsf{C}(A_k) \neq \mathsf{C}(A_h)}d(A_k, A_h) = \frac{1}{1 - \varepsilon},
\end{equation*}
with probability at least $1 - \delta$. The first part of the claim as in \eqref{eq:stronger_claim} holds.

In order to prove the second part of the claim, note that it holds
\begin{equation*}
\mathbb{P}\left (\mathsf{C}(A_{\mathrm{rand}} \right) = 1) = \frac{ 1}{\left \lceil 1/\delta \right \rceil} \qquad  \text{and}  \qquad \mathbb{P}\left(\mathsf{C}(A_{\mathrm{rand}}\right) = 1 - \varepsilon) = \frac{\left \lceil 1/\delta \right \rceil - 1}{\left \lceil 1/\delta \right \rceil}.
\end{equation*}
Hence, it holds
\begin{equation}
\label{eq:fgh}
\mathsf{C}_{\mathrm{rand}} = \mathbb{E}\left [{C}(A_{\mathrm{rand}}) \right ] = \frac{ 1}{\left \lceil 1/\delta \right \rceil} + (1 - \varepsilon) \frac{\left \lceil 1/\delta \right \rceil - 1}{\left \lceil 1/\delta \right \rceil}.
\end{equation}
Using the estimate of $\mathsf{C}_{\mathrm{rand}}$ as in \eqref{eq:fgh}, it follows that
\begin{equation*}
\lim_{\delta \to 0}\frac{\mathsf{C}_{\mathrm{max}}}{\mathsf{C}_{\mathrm{rand}}} = \lim_{\delta \to 0}\frac{1}{\frac{ 1}{\left \lceil 1/\delta \right \rceil} + (1 - \varepsilon) \frac{\left \lceil 1/\delta \right \rceil - 1}{\left \lceil 1/\delta \right \rceil}} = \frac{1}{1 - \varepsilon}.
\end{equation*}
Hence, \Eqref{eq:stronger_claim} holds and the claim follows.
\end{proof}

\section{Additional Details for Synthetic Simulations}

We provide the experimental setup details for various orchestration scenarios (c.f. \Cref{sec:understanding_orch}) in terms of the agent capabilities and the cost of orchestration.
For all experiments in this section, we orchestrate over a stream of $N=1000$ data points spread uniformly across $M=3$ regions and $K=4$ agents. The data is randomly shuffled once before the simulation.
Note that throughout this discussion, the precise values do not matter but only their relative magnitude.

We represent the \emph{true} (but unknown to the orchestrator) agent capabilities as a $K \times M$ matrix, such that each row represents the agent's capabilities across $M$ regions. Hence, for our simulations we represent these capabilities as a $4 \times 3$ matrix. 
Each element of the row represents the true probability of a correct answer in the corresponding region. 
To generate a stream of $1000$ data points, which in this case is merely a binary correctness label, we sample from a Bernoulli distribution with the probability parameter corresponding to each entry in the matrix. Each element of the stream is assigned to a region uniformly at random such that in the complete stream we see all regions roughly equally, though not necessarily in a fixed order due to the initial shuffle.

In addition, for some expertise profiles, we also note cost profiles used during simulation. In a similar manner, we represent the cost as an $K \times M$ matrix, where each element represents the mean cost associated with using the actor in a region. For the generation of the stream of data points, we sample from a Gaussian distribution with the cost mean and a standard deviation of $2$ to simulate variability in cost.

\paragraph{Approximately Invariant Expertise.} The agent capabilities matrix used for the simulations is,
\begin{align*}
\begin{bmatrix}
    0.350 & 0.336 & 0.314 \\
    0.339 & 0.338 & 0.323 \\
    0.349 & 0.322 & 0.329 \\
    0.331 & 0.311 & 0.357
\end{bmatrix}    
\end{align*}
Here, all agents have roughly similar capabilities across the regions.

\paragraph{Dominant Expertise.} The agent capabilities matrix used for the simulations is,
\begin{align*}
\begin{bmatrix}
    \mathbf{0.650} & \mathbf{0.852} & \mathbf{0.877} \\
    0.399 & 0.298 & 0.303 \\
    0.079 & 0.076 & 0.069 \\
    0.031 & 0.091 & 0.274
\end{bmatrix}    
\end{align*}
Here, we see that all agents are dominated across all regions by agent $A_1$.

\paragraph{Dominant Expertise with Misaligned Cost.} In addition to the dominant expertise profile above, we observe the misaligned cost behavior in simulations using the following matrix,
\begin{align*}
\begin{bmatrix}
    50.915 & 120.683 & 110.287 \\
    51.582 & 111.053 & \mathbf{1.412} \\
    45.006 & \mathbf{1.568} & 123.644 \\
    \mathbf{1.971} & 100.274 & 121.872
\end{bmatrix}    
\end{align*}
Here we see that in terms of the cost, agent $A_4$ dominates in region $R_1$, $A_3$ dominates in region $R_2$, and agent $A_2$ in region $R_3$. This cost profile is misaligned with the otherwise dominant agent $A_1$ in terms of the expertise profile, rendering orchestration less effective.

\paragraph{Varying Expertise.} The agent capabilities matrix used for the simulations is,
\begin{align*}
\begin{bmatrix}
    \mathbf{0.650} & 0.076 & 0.274 \\
    0.399 & 0.298 & 0.303 \\
    0.079 & \mathbf{0.852} & 0.069 \\
    0.031 & 0.091 & \mathbf{0.877}
\end{bmatrix}    
\end{align*}
Here, we construct a capabilities matrix such that agent $A_1$ is dominant only on region $R_1$, agent $A_3$ is dominant only on region $R_2$, and agent $A_4$ is dominant only on region $R_3$.

\paragraph{Varying Expertise and Misaligned Cost.} In addition to the varying expertise profile, we use the same misaligned cost profile presented earlier. As noted in \cref{sec:understanding_orch}, such a setting also hurts the appropriateness of orchestration.
These profiles are visualized for the appropriateness of orchestration in \cref{fig:profiles_approp}. We compute the expected cost of a random orchestrator $\mathsf{C}_{\mathrm{rand}}$, i.e. select a random actor at each step in the stream of data, using the average performance of $50$ simulations.

\section{Additional Details for Rogers' Paradox}

We provide details for \cref{sec:RP}. In our experimental setup, orchestration is enabled and each agent is forced to pick whichever option has the highest success probability. We introduce an additional constraint that only one of three AI systems is unavailable for each world state (randomly selected each time the world state changes). For the baseline, agents can only learn individually. If we let some agents learn socially and some individually, we recover Rogers' Paradox baseline.

We next provide low-level parameter settings describing the simulation environment, as modified from ~\citet{collins2025revisiting}. 
For our experiment, we ran 1000 agents in the environment with 4000 steps. The environment changed every step with probability 0.0001. The survival rate at the end of each step, if adapted is 0.93, and if not adapted, is 0.85. The environment replenished at the end of every time step proportionally to the surviving agents (with a mutation rate of 0.005). The mutation rate flips which learning strategy new agents inherit their AI bias from by adding noise sampled from $N(0,0,1)$. 

The cost of individual learning is 0.05. The success rate of individual learning is 0.66 times the individual penalty, where the individual penalty is sampled randomly from $N(\mu=1,\sigma=0.1)$ when each agent is initialized. 

In contrast, social learning is free with zero transmission error. However, for agent $i$, it is spatially constrained to be only from agents $[i-5,i+5]$ or any one of the three AI systems: one AI system learns (by taking the mean) from social learners, one from individual learners, and one from all learners.  AI bias is initialized as 1, and social learners can choose to learn from AI as opposed to other agents with probability $\frac{\mathrm{bias}}{1+\mathrm{bias}}$.



\section{Additional Orchestration User Studies}
\label{userstudy_app}
We now provide additional details on our user study in Section~\ref{sec:real_orchestration}. We consider questions from three topics of the \textbf{MMLU dataset:} elementary mathematics, high school mathematics, and college mathematics. These topic names align with those proposed in~\citet{Hendrycks2020MeasuringMM}. The selected topics span a broad range of mathematical expertise. To enhance readability, we dynamically adjust the frontend to accommodate varying question lengths, ensuring clarity. In total, we collect 300 questions with 100 selected from each topic. We provide screenshots of our interface in Figure~\ref{fig:variants_examples}.

Prior to considering orchestration, we conducted a \textbf{Pilot Study} with $20$ participants to establish a baseline for how users perform on this task. Each participant answered $60$ questions, $20$ per topic, independently without a decision aid. The average performance estimates across the population were $79.5\%$ for Elementary Mathematics, $54.5\%$ for High School Mathematics, and $43.5\%$ for College Mathematics. The average completion time per participant was $62$ minutes.

\subsection{Orchestration with Lock-In}
In the main text, we center our discussion on orchestration with \textbf{lock-in}: that is, if the user chooses to outsource to either agent, they are bound by the agent's response.
In this setup, participants are presented with 60 multiple-choice questions, evenly distributed across the three topics with 20 questions per topic. They are assigned to one of five possible batches, each containing 60 questions, presented in a randomized order. Participants are informed of the topic for each question, for example, whether it pertains to elementary mathematics. To encourage thoughtful engagement, a $10$-second delay is implemented before participants can submit their responses. 

We leverage TogetherAI's API to obtain LLM responses for each question using Meta Llama 3.3 70B. The region-wise accuracy is as follows: $0.62$ for elementary mathematics, $0.44$ for high school mathematics, and $0.51$ for college mathematics. Participants earn $10$ points for each correct answer they provide independently with no penalty for incorrect responses. However, outsourcing incurs a cost ($-7$ for human agent and $-3$ for AI agent) and thus the rewards vary: outsourcing to a Human Agent yields $+3$ points for a correct answer and $-7$ points for an incorrect answer while outsourcing to an AI Agent results in $+7$ points for a correct answer and $-3$ points for an incorrect answer. We encourage future work to change this cost structure based on context-specific considerations.

In the \textbf{Base Variant}, $20$ participants were allowed to freely outsource to either a human or AI agent. They received no guidance in their choices and could outsource as many times as they wanted across topics. Once a participant outsourced a question, they could not change their answer. On average, participants achieved $75\%$ in Elementary Mathematics, $51\%$ in High School Mathematics, and $48\%$ in College Mathematics. The study took approximately $50$ minutes per participant.

The \textbf{Orchestration Variant} involved $20$ participants who had the option to outsource their answers to either a human or AI agent. Unlike the base variant, participants were not left to make these decisions independently: instead, they received guidance through suggestions generated by our orchestration framework, using the empirical utility estimation method outlined in Section~\ref{sec:framework}. The recommendations provided were simple and direct: \textit{``You should outsource this problem to a human agent."}, \textit{``You should outsource this problem to the AI agent."}, or \textit{``You should attempt this problem by yourself."} 
We maintain a Beta-Binomial for every user's performance alone in each region: we initialize the Beta-Binomial by rounding the average pilot study performance. Specifically, we use the prior were $4/5$ for Elementary Mathematics, $3/5$ for High School Mathematics, and $2/5$ for Collage Mathematics. Note that we fix \( p(A_k|R_m) \) for the human agent and AI agent based on the pilot study performance and LLM accuracy respectively. Since we are outsourcing, we do not consider the responses from the human or AI agents to deviate (in the limit) from these estimates.

\begin{table}[htb]
    \centering
    \renewcommand{\arraystretch}{1.2}
    \setlength{\tabcolsep}{10pt}
    \begin{tabular}{lcccc}
        \toprule
        \textbf{Category} & \textbf{Unaided} & \textbf{Baseline} & \textbf{Orchestration} & \textbf{Constrained Orch.} \\
        \midrule
        Elementary Math & $79.75 \pm 3.80$ & $75.25 \pm 3.13$ & $81.00 \pm 2.37$ & $82.50 \pm 2.73$ \\
        High School Math & $54.50 \pm 4.67$ & $51.00 \pm 2.96$ & $55.50 \pm 3.61$ & $60.25 \pm 3.05$ \\
        College Math & $43.50 \pm 4.65$ & $48.00 \pm 3.31$ & $47.75 \pm 3.95$ & $52.75 \pm 2.91$ \\
        \midrule
        \textbf{Overall} & $59.25 \pm 4.39$ & $58.08 \pm 3.14$ & $61.42 \pm 3.38$ & $65.17 \pm 2.90$ \\
        \bottomrule
    \end{tabular}
    \caption{Region-wise \& overall performance percentages (\%) with standard errors across different lock-in variants. Orchestration with constraints is most effective over all categories, but most effective in regions where users struggle (i.e., College Mathematics).}
    \label{tab:lockin-performance}
\end{table}

As participants answered questions unaided, their Beta-Binomial posterior estimates were updated. For example, if a participant answered correctly without outsourcing, the estimate was updated using the formula \( \frac{4+1}{5+1} \). Conversely, for an incorrect answer, the update followed the formula \( \frac{4}{5+1} \). However, if a participant chose to outsource a question, no Beta-Binomial update was applied. For each new question, the orchestration framework compared empirical utility estimates \( \widehat{\mathsf{U}}_{\geq t}(A_k) \) and recommended the agent with the highest utility. This ensures that suggestions adapted in real-time, taking into account the participant’s ongoing performance and the relative performance of the human and AI agents. Participants performed notably well under this setup, achieving roughly $81\%$ accuracy in Elementary Mathematics, $56\%$ in High School Mathematics, and $48\%$ in College Mathematics. On average, participants took $42$ minutes to complete the study.

The \textbf{Constrained Orchestration Variant}, described in Section~\ref{sec:framework}, builds upon the same orchestration framework but introduced an additional rule to guide decision-making: if a participant answered a question incorrectly without outsourcing, they were required to outsource their next question in the same region. In such cases, the option to attempt the next question independently was disabled. For these subsequent questions, the orchestration framework restricted the empirical utility comparison to just the (fixed) AI and human agents. 
To make this constraint clear, participants received a direct message reinforcing the rule, such as: \textit{``You were wrong by yourself on " + region + `` last time. You should outsource this problem to a human agent."} or \textit{``You were wrong by yourself on " + region + `` last time. You should outsource this problem to the AI agent."}. This strikes a balance between autonomy and outsourcing. The results reflected a notable improvement in performance, with  $20$ participants achieving about $83\%$ accuracy in Elementary Mathematics, $60\%$ in High School Mathematics, and $53\%$ in College Mathematics. On average, participants took $35$ minutes to complete this variant.

\begin{figure}[htb]
    \centering
    \includegraphics[width=0.7\linewidth]{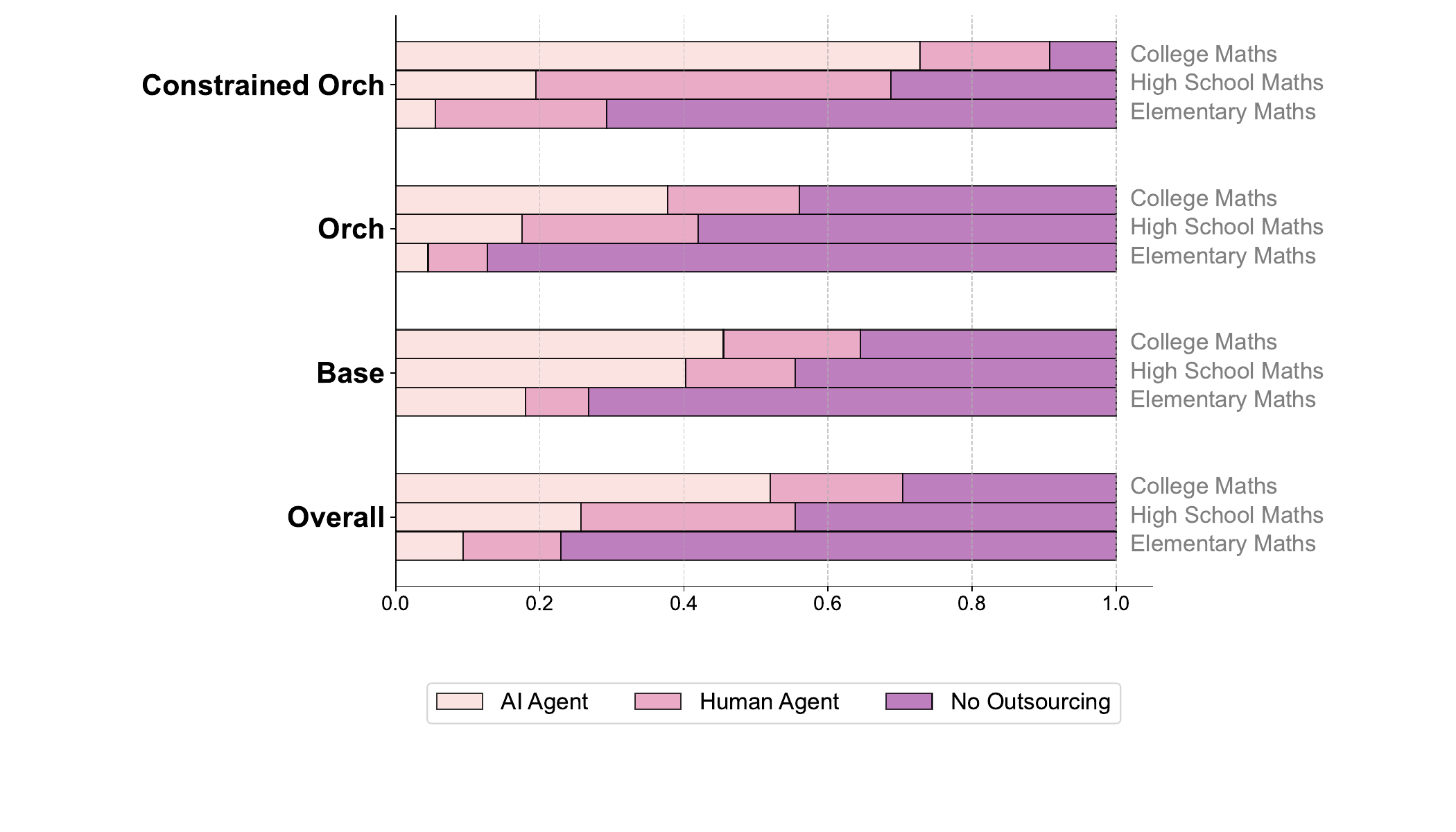}
    \vspace{-1cm}
    \caption{The proportion of questions outsourced to human and AI agents, as well as those answered independently by participants, across the different regions of lock-in variants — including base, orchestration, and constrained orchestration.}
    \label{fig:regional_outsourcing_variants_lock-in}
\end{figure}

In Table~\ref{tab:lockin-performance}, we summarize the performance across regions and lock-in variants. In addition, Figure~\ref{fig:overall_performance} shows how constrained orchestration outperforms orchestration in all regions. 
In Figure~\ref{fig:regional_outsourcing_variants_lock-in}, we show how the rate of outsourcing to AI agents differs across regions significantly between the three variants; 
for example, outsourcing to AI agents is very common for college mathematics under constrained orchestration, likely due to users being forced to outsource after trying those problems alone. 
A similar trend is observed in human outsourcing, with statistically significant differences across variants in all regions.


\subsection{Orchestration with No Lock-In}
For the sake of completeness, we consider orchestrating the use of agents when users have the flexibility to override an agent's suggestion. 
Unlike the main-text study in which users are bound to the agent's response upon outsourcing, users retain control in the non-lock-in study as they have the freedom to review and modify the recommendation.

\begin{table}[htb]
    \centering
    \renewcommand{\arraystretch}{1.2} 
    \setlength{\tabcolsep}{10pt} 
    \begin{tabular}{lcccc}
        \toprule
        \textbf{Category} & \textbf{Unaided} & \textbf{Baseline} & \textbf{Orchestration} & \textbf{Constrained Orch.} \\
        \midrule
        Elementary Math & $79.75 \pm 3.80$ & $81.25 \pm 3.85$ & $81.75 \pm 3.89$ & $87.00 \pm 2.12$ \\
        High School Math & $54.50 \pm 4.67$ & $52.50 \pm 4.03$ & $54.50 \pm 3.84$ & $61.50 \pm 3.65$ \\
        College Math & $43.50 \pm 4.65$ & $51.00 \pm 3.60$ & $48.25 \pm 3.69$ & $50.50 \pm 2.59$ \\
        \midrule
        \textbf{Overall} & $59.25 \pm 4.39$ & $61.58 \pm 3.83$ & $61.50 \pm 3.81$ & $66.33 \pm 2.86$ \\
        \bottomrule
    \end{tabular}
    \caption{Region wise \& overall performance percentages (\%) with standard errors across different non-lock-in variants.}
    \label{tab:Nonlockin-performance}
\end{table}

The same pilot study as before holds, as there are no agents present. In the base variant, users can select to see the response of either agent (human or AI) but can choose to override the suggestion. On average, participants achieved $81\%$ accuracy in Elementary Mathematics, $53\%$ in High School Mathematics, and $51\%$ in College Mathematics. The study took approximately $44$ minutes per participant to complete. The base variant without lock-in outperforms the lock-in variant across all regions. This stems from the fact that, in the lock-in variant, users often outsource and stick to the agent’s incorrect responses. In contrast, without lock-in, users can override the agent's mistakes, leading to better overall accuracy.

In the \textbf{Orchestration Variant}, we again use empirical utility estimates derived from pilot study performance and LLM accuracy to initialize  a prior for each region. In this study, we initialize a Beta-Binonimal for every actor-region pair: we do not assume that the performance of the user with an agent is static, as we did in the main-text study. This change is natural since, without lock-in, user performance after seeing an agent's response may vary from the agent's performance due to overriding~\cite{guo2024decision}.  Listed in the order [Human Agent, AI Agent, User], the priors for Elementary Mathematics were $[40/50, 31/50, 4/5]$, for High School Mathematics $[27/50, 22/50, 3/5]$, and for College Mathematics $[22/50, 26/50, 2/5]$. 
Whenever a participant answered a question, whether independently or with outsourcing, a Beta-Binomial estimate was updated. The prior for agent use (human and AI) was ten times stronger compared to that of the user: we assume that, to start, the effect of an user's error alone is larger than the effect of a user's error from agent use.  As a result, participants achieved, on average, $82\%$ in Elementary Mathematics, $55\%$ in High School Mathematics, and $48\%$ in College Mathematics. On average, the study took approximately $38$ minutes per participant to complete.

We run the same \textbf{Constrained Orchestration Variant} as with the lock-in study. If a participant answered a question incorrectly by themselves, they were required to outsource their next question within that same region. Though outsourcing was mandated, users were still able to override the agent's response. We find that participants achieved approximately $87\%$ in Elementary Mathematics, $62\%$ in High School Mathematics, and $51\%$ in College Mathematics. The average time to complete this study was $63$ minutes per participant.

\begin{figure}[!t]
    \centering
    \includegraphics[width=0.7\linewidth]{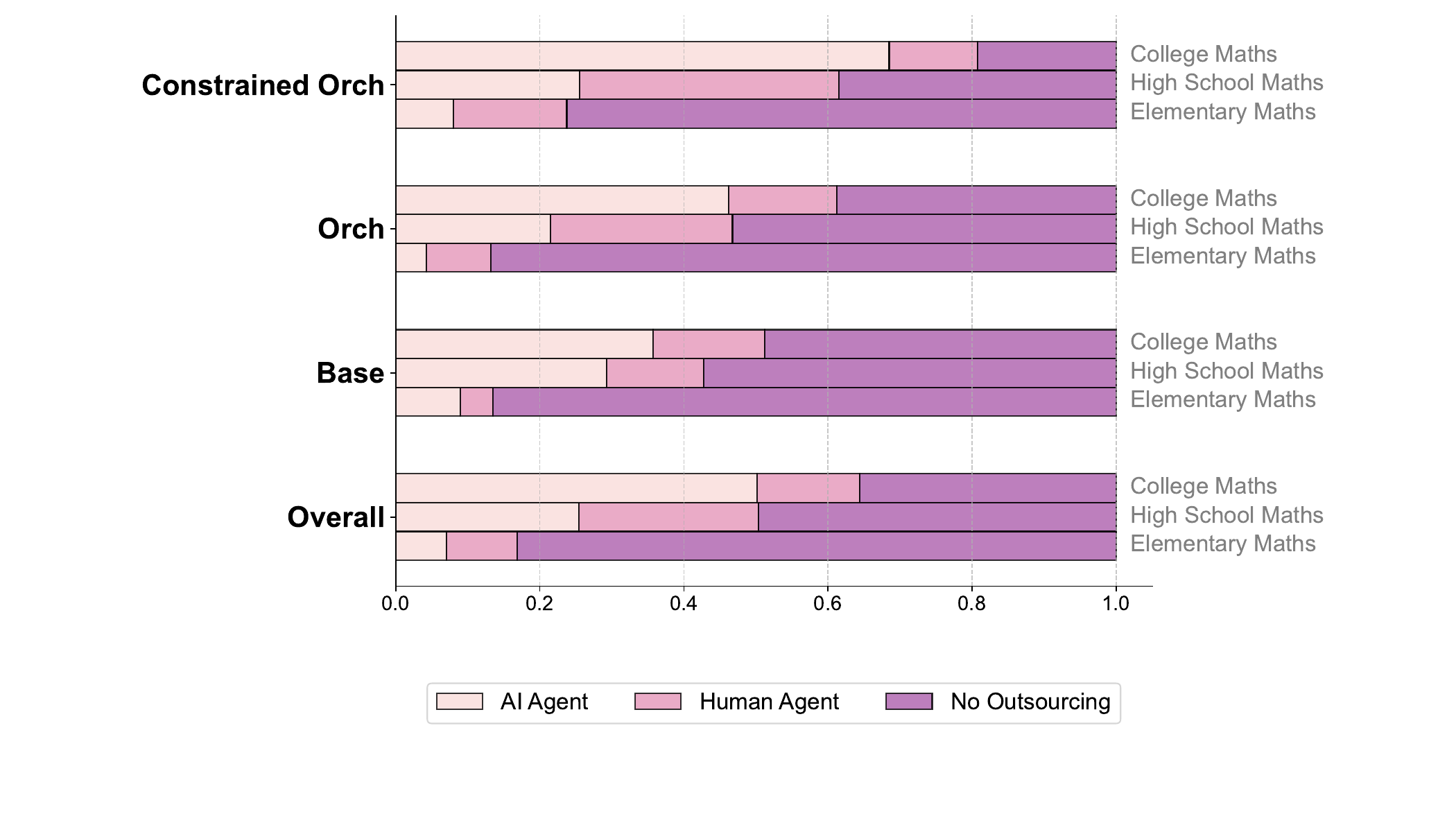}
    \vspace{-1cm}
    \caption{The proportion of questions outsourced to human and AI agents, as well as those answered independently by participants, across the different regions of non-lock-in variants — including base, orchestration, and constrained orchestration.}
    \label{fig:regional_outsourcing_variants_non-lock-in}
\end{figure}

\begin{figure}[!t]
    \centering
    \includegraphics[width=0.7\linewidth]{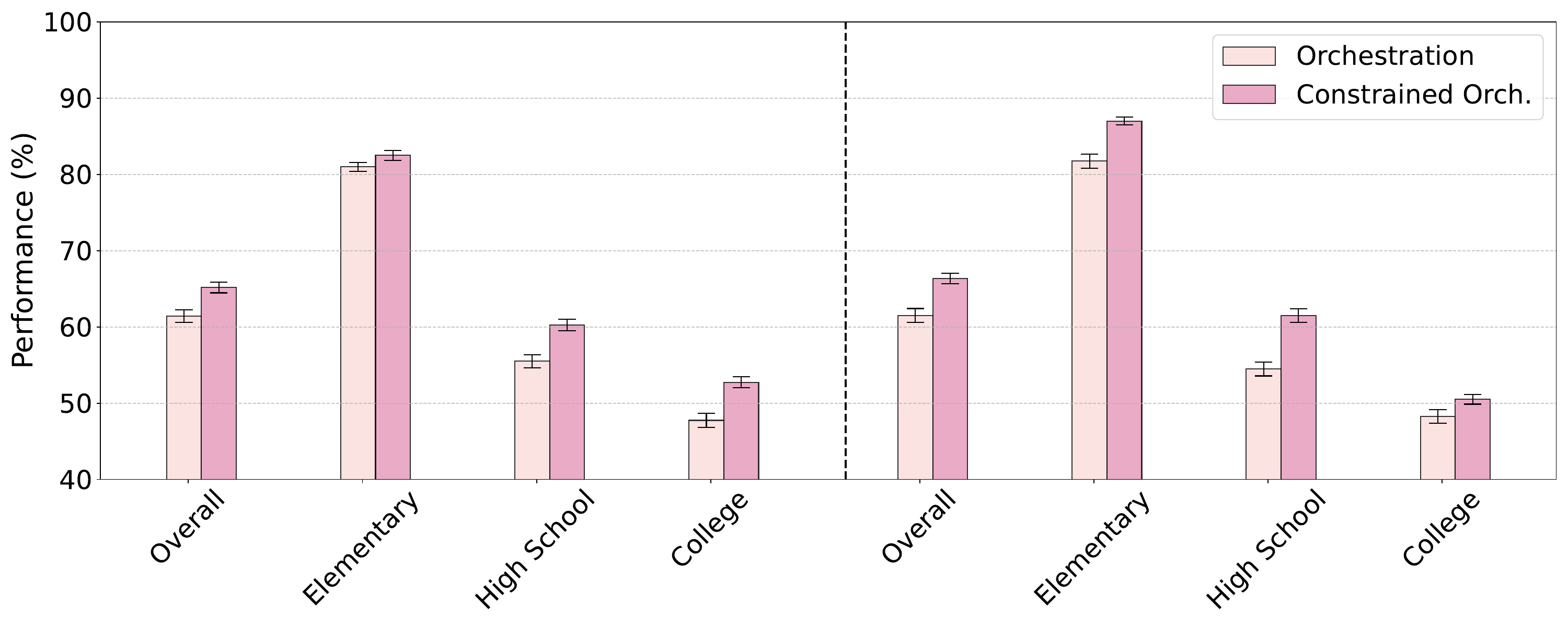}
    \caption{The figure compares overall and regional accuracy across lock-in and non-lock-in variants of the study, evaluating both orchestration and constrained orchestration. The left side illustrates results for the lock-in study, while the right side presents findings for the non-lock-in study. Constrained orchestration consistently enhances performance across all regions in both studies. Notably, regional performance in the lock-in study is generally lower compared to their non-lock-in counterparts.
    }
    \label{fig:overall_performance}
\end{figure}

All of these results are detailed in Table~\ref{tab:Nonlockin-performance}, we summarize the performance across regions and non-lock-in variants. In Figure~\ref{fig:overall_performance}, we visualize how constrained orchestration outperforms orchestration across various regions of the non-lock-in variant.
As shown in Figure~\ref{fig:regional_outsourcing_variants_non-lock-in}, outsourcing rates differ across regions and variants.


\subsection{Further Results}

\textbf{Participant Comments Hint at Automation Bias and Challenges to User Autonomy} In a post-experiment survey where participants were invited to share general feedback, we present a selection of comments that we found particularly revealing.

\begin{itemize}
    \item \textit{``I didn't like that there were some questions where I knew the answer, but I was forced to use either AI or human outsourcing for the answer.''}
    
    \item \textit{``However, I didn't like being forced to use a human/AI agent in some cases.''}
    
    \item \textit{``There were some arcane topics that I didn't even bother with. I usually outsourced these.''}
    
    \item \textit{``I notice that solving the questions myself would take some time, I just end up using the AI or human.''}
\end{itemize}

We observed recurring themes around the tension between individual problem-solving and the imposed use of AI or human agents. Several participants expressed discomfort with being required to outsource answers, even in cases where they felt confident in their own knowledge. Moreover, participants also highlighted a pattern of selectively outsourcing difficult or time-consuming questions, often defaulting to AI or human help for questions topics without even trying. We believe that these responses highlight the need for further exploration into how mandatory AI or human assistance influences user confidence, autonomy, and problem-solving. Understanding these dynamics is essential to mitigate potential automation bias and find a balance between user control and external support, particularly in decision-making tasks \cite{kupfer2023check}.

\begin{figure}[H]
    \centering
    \begin{subfigure}{0.48\textwidth}
        \includegraphics[width=\linewidth]{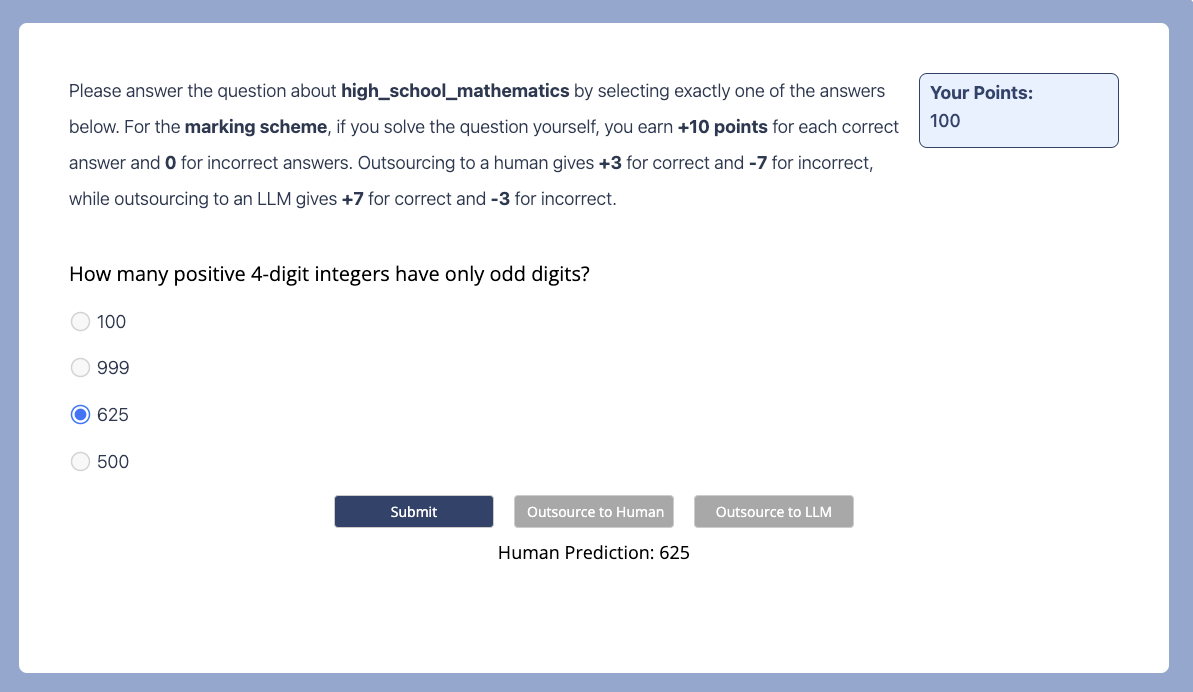}
    \end{subfigure}
    \hspace{0.02\textwidth}
    \begin{subfigure}{0.48\textwidth}
        \includegraphics[width=\linewidth]{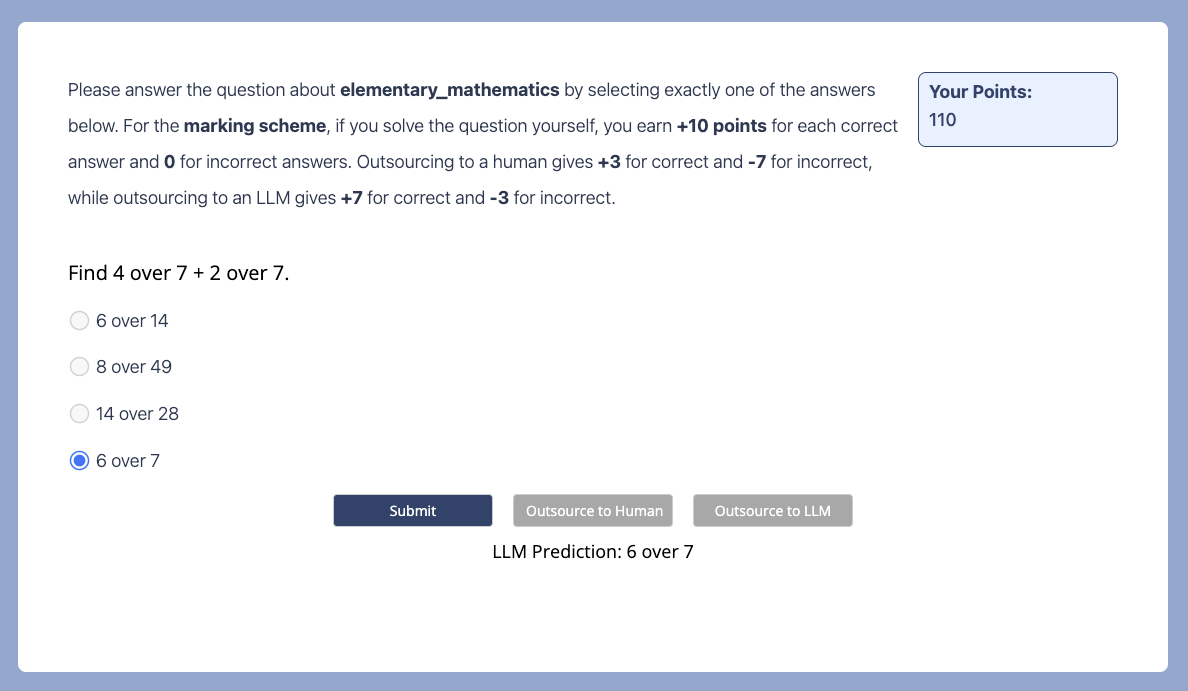}
    \end{subfigure}
    
    \begin{subfigure}{0.48\textwidth}
        \includegraphics[width=\linewidth]{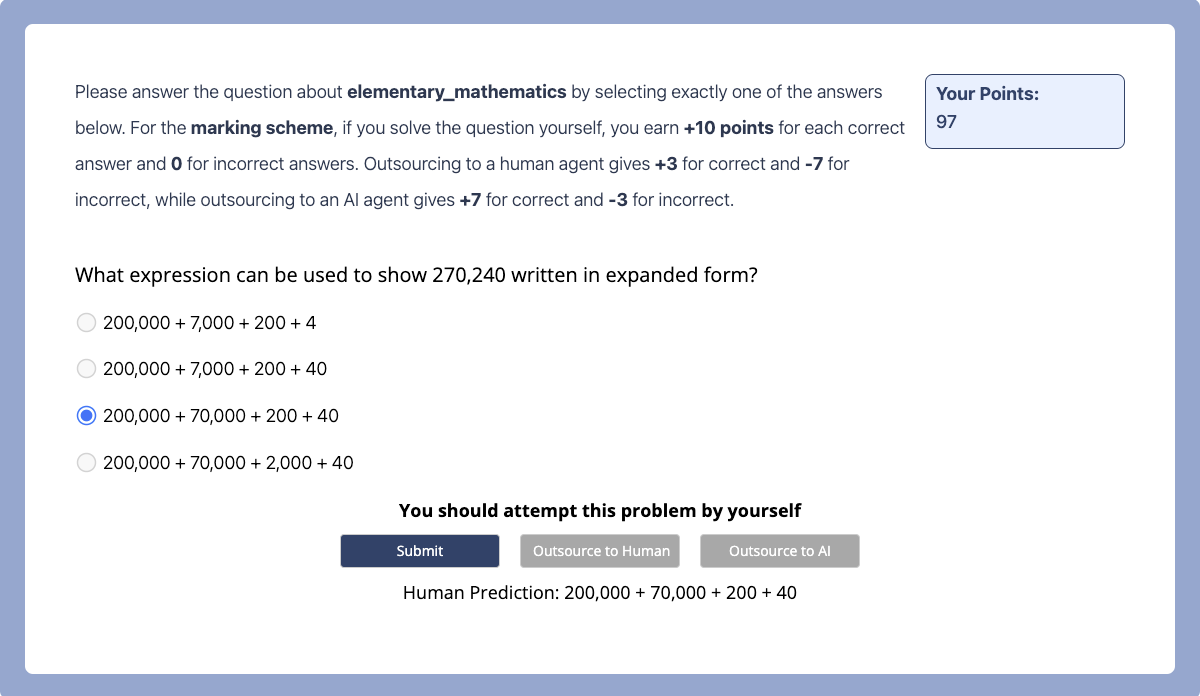}
    \end{subfigure}
    \hspace{0.02\textwidth}
    \begin{subfigure}{0.48\textwidth}
        \includegraphics[width=\linewidth]{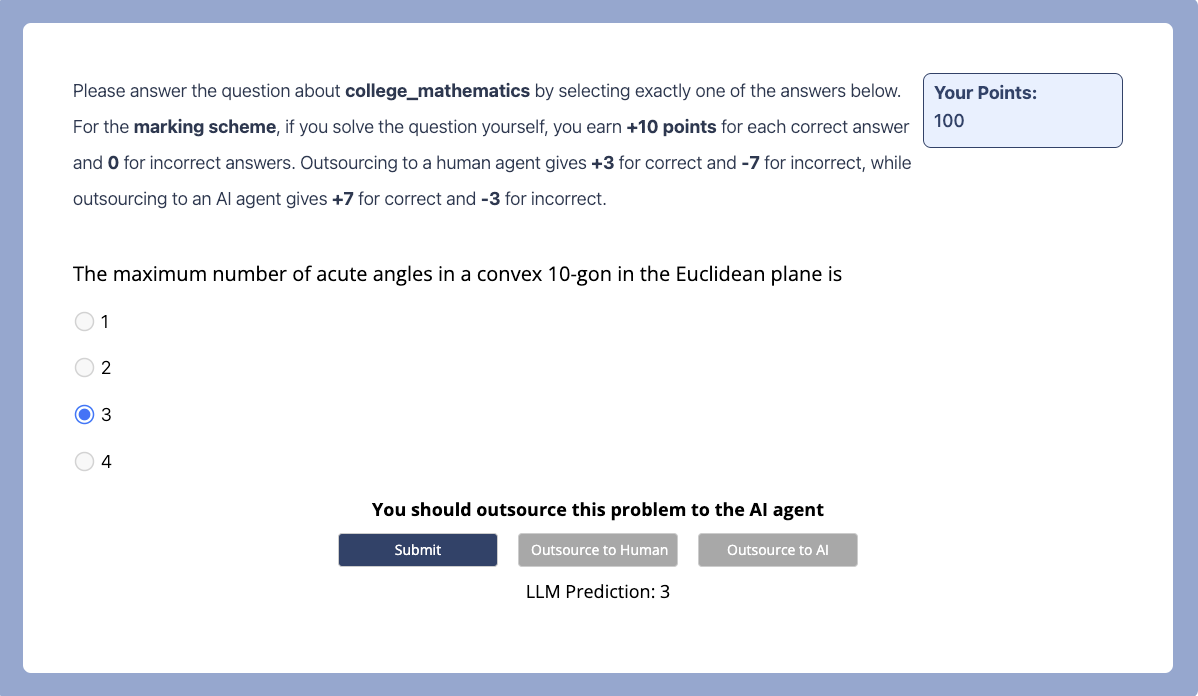}
    \end{subfigure}
    
    \begin{subfigure}{0.48\textwidth}
        \includegraphics[width=\linewidth]{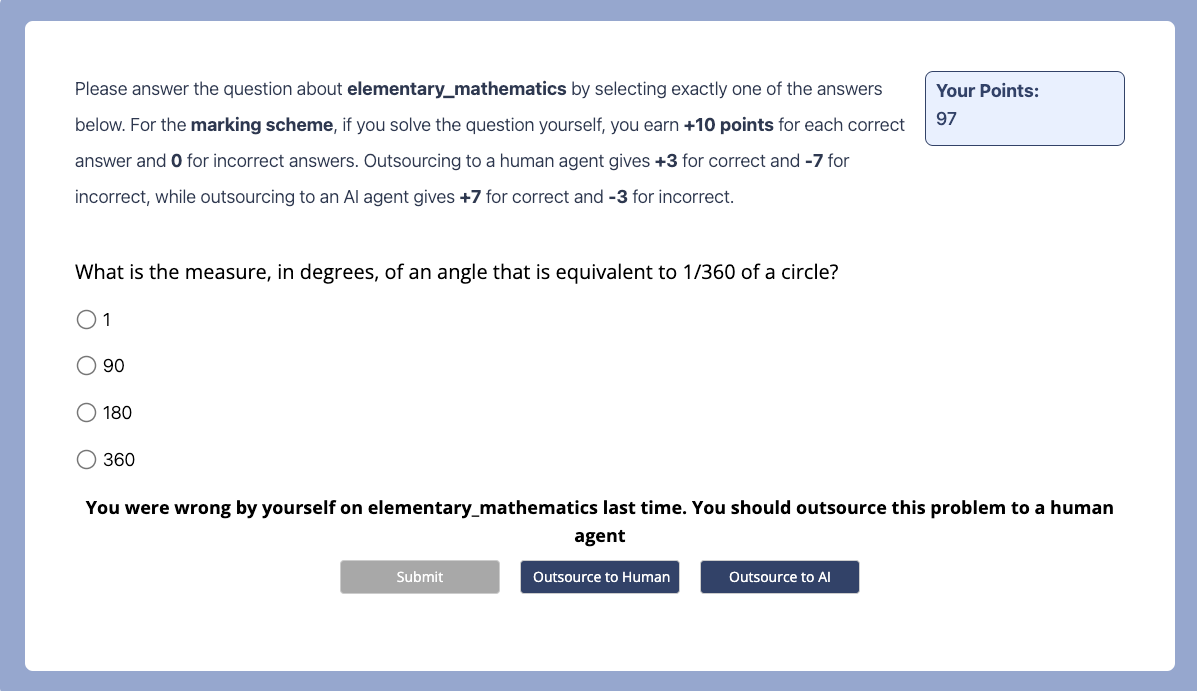}
    \end{subfigure}
    \hspace{0.02\textwidth}
    \begin{subfigure}{0.48\textwidth}
        \includegraphics[width=\linewidth]{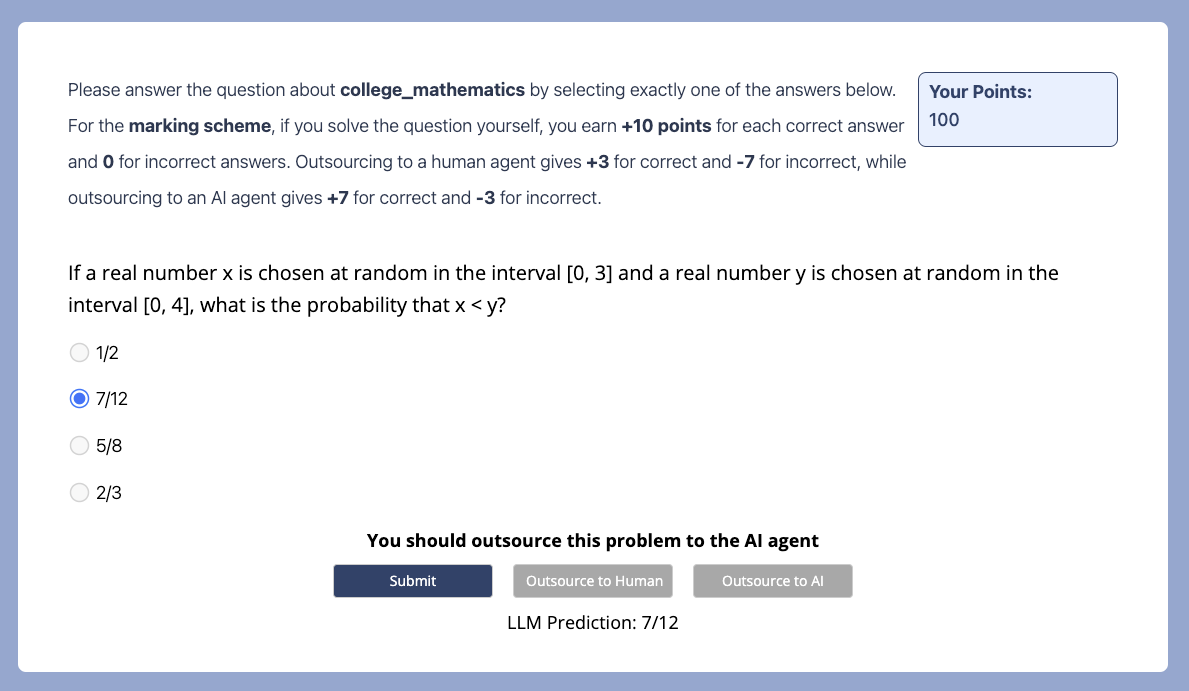}
    \end{subfigure}

    \caption{Interface examples of the lock-in Baseline (Top), Orchestration (Middle), and Constraint Orchestration (Bottom) variants.}
    \label{fig:variants_examples}
\end{figure}

In Figure~\ref{fig:variants_examples}, we depict examples of our user interface.

\end{document}

%% file: math_commands.tex
\usepackage{amsmath,amsfonts,bm}

\def\eqref#1{equation~\ref{#1}}
\def\Eqref#1{Equation~\ref{#1}}

\def\1{\bm{1}}

\DeclareMathAlphabet{\mathsfit}{\encodingdefault}{\sfdefault}{m}{sl}
\SetMathAlphabet{\mathsfit}{bold}{\encodingdefault}{\sfdefault}{bx}{n}

\def\gA{{\mathcal{A}}}

\def\gX{{\mathcal{X}}}

